\documentclass[fleqn,usenatbib,dvipdfmx]{mnras}
\usepackage{lineno}
\DeclareRobustCommand{\VAN}[3]{#2}
\let\VANthebibliography\thebibliography
\def\thebibliography{\DeclareRobustCommand{\VAN}[3]{##3}\VANthebibliography}
\newcommand{\oneE}{1E~1547.0$-$5408}    
\newcommand{\NuS}{{\it NuSTAR}}
\newcommand{\ASCA}{{\it ASCA}}
\newcommand{\Su}{{\it Suzaku}}
\newcommand{\zz}{Z_4^2}
\newcommand{\Tppm}{T_{\rm ppm}}
\newcommand{\EL}{E_{\rm L}}
\newcommand{\EU}{E_{\rm U}}
\newcommand{\Es}{E_{\rm s}}
\newcommand{\Ee}{E_{\rm e}}
\newcommand{\Ppr}{P_{\rm pr}}
\newcommand{\Prot}{P_{\rm rot}}
\newcommand{\tauc}{\tau_{\rm c}}
\newcommand{\Pch}{{\cal P}_{\rm ch}}
\newcommand{\SGR}{SGR 0501+4516}
\newcommand{\RXJ}{RXJ~1708$-$4009}
\newcommand{\Kes}{1E~1841$-$045}
\newcommand{\Bd}{B_{\rm d}}
\newcommand{\Bt}{B_{\rm t}}
\usepackage{graphicx}	
\usepackage{amsmath}	
\usepackage{amssymb}
\usepackage{booktabs}

\title{Observational Clues to the Magnetic Evolution of Magnetars }

\author[K. Makishima et al.]{
Kazuo Makishima,$^{1}$
\thanks{E-mail: maxima@phys.s.u-tokyo.ac.jp (KM)}
Nagomi Uchida,$^{2}$
and
Teruaki Enoto$^{3}$
\\
$^{1}$Department of Physics, The University of Tokyo,
7-3-1 Hongo, Bunkyo-ku, Tokyo, Japan 113-0033\\
$^{2}$ Institute of Space and Astronautical Science, JAXA,
3-1-1 Yoshinodai, Chuo-ku, Sagamihara 252-5210, Japan\\
$^{3}$ Department of Physics, Kyoto University,
Kitashirakawa Oiwake-cho, Sakyo-ku, Kyoto, Japan 606-8502 \\
}
\date{Accepted XXX. Received YYY; in original form ZZZ}
\pubyear{2024}
\begin{document}
\label{firstpage}
\pagerange{\pageref{firstpage}--\pageref{lastpage}}
\maketitle
\begin{abstract}
Utilizing four archival X-ray datasets  taken with the Hard X-ray Detector onboard \Su,
timing studies were performed on three magnetars, 
\Kes\ (observed in 2006), \SGR\ (2008), and 1RXS~J170849.0$-$400910 (2009 and 2010).
Their pulsations were reconfirmed,
typically in an energy range of 12--50 keV.
The $11.783$ s  pulses of \Kes\ 
and those of \SGR\  with $5.762$ s were  periodically phase modulated,
with a long period of $\approx 23.4$ ks and $\approx 16.4$ ks, respectively.
The  pulse-phase modulation
was also observed, at  $\approx 46.5$ ks, 
from two datasets of 1RXS~J170849.0$-$400910.
In all these cases, the modulation amplitude was 6 to 16\% of the pulse cycle.
Including previously confirmed four objects,
this characteristic timing behavior is now detected from seven magnetars in total,
and  interpreted as a result of free precession
of neutron stars that are deformed to an asphericity of $\sim 10^{-4}$.
Assuming that the deformation is due to magnetic stress,
these magnetars are inferred to harbor toroidal magnetic fields of $\Bt \sim 10^{16}$ G.
By comparing the estimated $\Bt$ of these objects with their  poloidal
dipole field $\Bd$, the $\Bt/\Bd$ ratio is found to increase with their characteristic age.
Therefore, the toroidal fields of magnetars are likely to last
longer than their dipole fields.
This explains the presence of some classes of  neutron stars
that have relatively weak $\Bd$
but are suspected to hide strong $\Bt$ inside them.
\end{abstract}


\section{INTRODUCTION}
\label{sec:intro}

Outstanding characteristics of neutron stars (NSs)
include their extremely intense magnetic fields.
Nevertheless, it remains still unknown
how these magnetic fields are produced, sustained, and dissipated.
Toward elucidation of these issues 
of fundamental astrophysical importance,
we expect valuable clues to be provided by magnetars 
\citep{Mereghetti08, Kaspi17, EnotoKisakaShibata},
i.e., an extreme  subclass of NSs with the strongest magnetic fields.
Indeed, magnetars are thought to have 
poloidal dipole magnetic fields  reaching $\Bd \lesssim 10^{15}$ G,
and emit mostly in X-rays by dissipating their magnetic energies.
Furthermore, magnetars may not be rare objects,
and can in fact dominate the new-born NSs,  
as suggested by \citet{Nakano15} based on X-ray observations
and a modeling of magnetic-field decay.

Magnetars, in particular those  called  Soft Gamma Repeaters  (SGRs), 
often enter  an active state,  
wherein they rapidly repeat the emission of ``soft gamma-ray bursts",
each lasting for $\lesssim 0.5$ s,
which can sometimes reach us even from external galaxies \citep{Mereghetti24}.
These short bursts must be an essential activity mode of magnetars,
because their persistent X-ray emission  could be 
an accumulation of numerous ``microbursts" \citep{Nakagawa11}.
Furthermore, the burst activity of magnetars could sometimes 
be  accompanied by the enigmatic phenomenon known as Fast Radio Bursts (FRBs)
 \citep{FRBMGT1,FRBMGT2, Mereghetti20, FRBMGT3, ULPM}.
These activities  are considered to result from 
a rapid release of magnetic energies,
although it is still unclear how and where the release takes place.

One  important aspect of  magnetars is their toroidal magnetic field, $\Bt$, 
as several observations suggest that $\Bt$ can be higher,
and last longer, than their $\Bd$.
For example, some NSs such as ``weak-field magnetars"
with $\Bd \approx {\rm several~times} ~10^{13}$ G
(e.g., \citealt{WeakFieldMGT10, Turolla+Esposito13,WeakFieldMGTs16}),
some CCOs (Central Compact Objects) in 
supernova remnants (e.g., \citealt{Rea16}), 
and ultra-long-period radio pulsars (e.g., \citealt{ULPM}),
can be magnetar-like objects with high $\Bt$ but low $\Bd$.
In addition, a dipole magnetic field  of $\Bt \sim 10^{14}$ G can sustain
the X-ray luminosity of a typical magnetar,
$\sim 3 \times 10^{34}$ erg s$^{-1}$, only for $\sim 5$ kyr.
Since this is not long enough for their characteristic age 
$\tauc \equiv P/2\dot{P}$ (with $P$ the pulse period and $\dot P$ its change rate),
their energy release may  come, at least partially, from $\Bt$ as well.
Also from theoretical view points, a magnetar could harbor  $\Bt\sim 10^{16}$ G  
(e.g., \citealt{Braithwaite,Fujisawa23}),
which is perhaps created during its formation, 
as the progenitor core collapses involving rapid differential rotation 
(e.g., \citealt{toroidal_B}).

While $\Bd$ of a NS deforms it  into an oblate shape,
$\Bt$ causes a prolate deformation
\citep{Cutler02,Ioka&Sasaki04}.
Let  $I_j~ (j=1,2,3)$ be the star's principal moments of inertia
in the coordinates $(\hat{x}_1,\hat{x}_2,\hat{x}_3)$ fixed to it,
with $\hat{x}_3$ the  symmetry axis.
When $\Bt \gg \Bd$, the expected  asphericity will be
\begin{equation}
\epsilon  \equiv (I_1 - I_3)/I_3\sim 10^{-4} C (B_{\rm t}/10^{16}{\rm G})^2
\label{eq:Epsilon_Bt}
\end{equation}
\citep{Cutler02},  where $C$ is a numerical factor of order unity.
This $\epsilon$ represents the ratio between 
the magnetic and gravitational energies of the star,
and will provide nearly the sole mean
to detect $\Bt$  hidden inside NSs.

When a torque-free rigid body
with a constant angular momentum $\vec L$ is  axially  deformed to  have $\epsilon \ne 0$,
its  motion around the center of gravity is  described 
as a superposition of two modes \citep{Landau&Lifshitz, Butikov06};
the  rotation around $\hat{x}_3$ 
with a rotation period $\Prot=2 \pi I_3/| \vec L| $,
and  the free precession in which $\hat{x}_3$ rotates around $\vec L$,
keeping a constant wobbling angle $\alpha \ne 0$,
with  a precession period $\Ppr = 2 \pi I_1/| \vec L| = (1+\epsilon) \Prot$.
When the body's emission is symmetric around $\hat{x}_3$,
we can detect only $\Ppr$ as the pulsation.
If the emission breaks the symmetry around $\hat{x}_3$,
we observe a mixture of the two periods;
the  pulsation with a period  $\Ppr$ becomes phase-modulated
at the beat period between $\Ppr$ and $P_{\rm rot}$, 
which is given as
\begin{equation}
\Tppm = \frac{\Ppr}{\epsilon \cos \alpha} 
= \frac{1}{\cos \alpha} \left(P_{\rm rot}^{-1} -{\Ppr}^{-1} \right)^{-1}~.
\label{eq:slip_period}
\end{equation}
This effect is named  {\it pulse-phase modulation} (PPM),
and $\Tppm$ is called {\em slip period.}
If PPM is detected,
we can estimate the deformation as $\epsilon \sim \Ppr/\Tppm$, assuming $\cos \alpha \sim 1$.
Sometimes, $\Tppm$ itself, instead of $\Ppr$, is called {\em precession period} 
(e.g, \citealt{Gao23, Zanazzi20}),
but this difference is simply a matter of nomenclature.

Using archival X-ray data from \Su, \NuS, and {\em ASCA},
we have so far observed firm evidence of the PPM from four magnetars,
4U 0142+61 \citep{Makishima14,Makishima19}, 
\oneE\  \citep{Makishima23, Makishima21a}, 
SGR 1900+14 \citep{Makishima21b}, and SGR 1806$-$20 \citep{Makishima24}.
In all of them, the PPM  resides in their  hard X-ray component,
which dominates the spectrum above $\sim 10$ keV as a power-law shaped hard tail.
Moreover, the four sources are all inferred to have 
$\epsilon \sim 10^{-4}$  from Equation~(\ref{eq:slip_period}),
and hence $\Bt \sim 10^{16}$ G via Equation~(\ref{eq:Epsilon_Bt}).

The above studies also gave an indication 
that magnetars with larger $\tauc$  have higher $\Bt/\Bd$ ratios \citep{Makishima24}.
However, the four objects are either rather aged ($\tauc \approx 68$ ks; 4U~0142+61)
or very young ($\tauc < 1$ kyr; the other three),
with no information on sources with intermediate $\tauc$.
To fill the age gap, in the present study we search for the PPM effects in three magnetars,
SGR~0501+4516, 1RXS~J170849.0$-$400910 (abbreviated as \RXJ),
and 	1E~1841$-$045,
which have  $\tauc=15$ kyr, 8.9 kyr,
and 4.6 kyr, respectively  \citep{McGill}.
On \SGR\ and \RXJ, preliminary results were already 
obtained affirmatively \citep{Makishima23}.

\section{OBSERVATIONS}
\label{sec:obs}
As summarized in Table~\ref{tbl:obs} and described below,
we utilize four archival datasets  of the three objects, 
taken with the fifth Japanese X-ray satellite \Su.
We  analyze the data from the HXD-PIN detector \citep{HXD1, HXD2} onboard \Su,
but not those from HXD-GSO because it  detected  no magnetar signals.
The soft X-ray data from the {\Su}'s  X-ray Imaging Spectrometer \citep{XIS}
are not analyzed, either, 
because its data usually have insufficient time resolution (either 2 s or 1 s),
and our major interest is the hard component signal
which is mostly outside the XIS energy coverage
(typically 0.3--10 keV).

\begin{table*}
\caption{\Su\ HXD datasets utilized in the present work.}
\label{tbl:obs}
\begin{center}
\begin{tabular}{lcccccccccccc}
\hline
Object & ObsID& Start date  &MJD $^a$ &  Data span$^b$&  Exposure$^c$ &Total photons$^d$ & Flux$^e$ \\
\hline
\Kes\ & 401100010 &2006 April 19 & 53844.45  &  229.2 ks  & 59.8 ks & 40014  &  $48.9\pm 0.3$ \\
\hline
\SGR\ & 903002010 &2008 August 26 &54704.02 &  110.8 ks  & 50.7 ks &  27271 &  $28.1 \pm 6.5$  \\
\hline
\RXJ & 404080010 &2009 August 23      &  55066.68 & 108.5 ks  & 47.9 ks & 27682 &  $24.4 \pm 4.4$   \\
\RXJ & 405076010& 2010 Sept. 27 &  55466.61    &   99.1   ks  & 55.4 ks & 31280 &$ 24.4 \pm 4.0$   \\
\hline

\end{tabular}
\begin{footnotesize}
\begin{itemize}
\setlength{\itemsep}{0mm}
\item[$^a$] The Modified Julian Date of the first photon in the data.
\item[$^b$] Total elapsed time of the observation, from the start to the end.
\item[$^c$] The net exposure by the HXD, taken from \citet{Enoto17}.
\item[$^d$]  The total number of 10--70 keV events detected with  HXD-PIN, including background.
\item[$^e$]  Absorption-inclusive 15--60 keV flux detected with HXD-PIN,
with $1\sigma$ errors, in $10^{-12}$ erg s$^{-1}$ cm$^{-2}$. From \citet{Enoto17}.
\end{itemize}
\end{footnotesize}
\end{center}
\end{table*}

\subsection{1E~1841$-$045}
\label{subsec:obs_1841}
The Anomalous X-ray Pulsar (AXP) \Kes\ resides at the projected center 
of the supernova remnant Kes 73 \citep{Vasisht97}.
As given in Table~\ref{tbl:obs}, it was observed  in 2006 with \Su\ for a gross exposure of 229 ks.
The data were already analyzed by \citet{Morii10},
who detected the pulsation with the XIS and HXD-PIN,
at a period of $P=11.7830(2)$ s.
In addition, these authors detected the hard X-ray component  significantly with HXD-PIN,
over a typical energy range of 15--40 keV.
We are hence encouraged to analyze the same 
HXD-PIN data for the PPM effects.

\subsection{SGR~0501+4516}
\label{subsec:obs_0501}

On 2008 August 22,  many short X-ray bursts were detected 
with the {\em Swift} Burst Alert Telescope,
from a new SGR source,
which was then named  \SGR\ \citep{GCN1, GCN2}.
On this alert, a Target-of-Opportunity observation of the source
was conducted with \Su\ for a gross pointing of $\approx 111$ ks.
Using the  acquired data, 
a few papers were already published.
In \citet{Enoto09}, the HXD and XIS data were analyzed
for the pulsation,  spectra, and  short bursts,
focusing on the soft X-ray component.
The pulsation, which had been discovered several days 
before with {\it RXTE} \citep{Gogus08},
was reconfirmed at a period of $P=5.762072(2)$ s.
Then, \citet{Enoto10b} achieved a detection of the hard X-ray component
in the persistent spectrum  up to $\sim 60$ keV;
this makes \SGR\ a promising candidate for the present study.
{Finally, \cite{Nakagawa11} accumulated
a 1--40 keV spectrum over bright bursts,
and found that it also consists of  the soft and hard components,
just like the persistent spectrum.

\subsection{\RXJ\ (1RXS J170849.0$-$400910)}
\label{subsec:obs_1708}
This X-ray source was first discovered with {\em ROSAT} \citep{Voges98}.
Later, \citet{Sugizaki97} detected a 11-s periodicity in its {\it ASCA} data,
and classified \RXJ\ as an AXP.
It also became one of the prototypical magnetars
from which the hard X-ray component was first detected with {\em INTEGRAL} \citep{Kuiper06}.
Therefore, this object is another good candidate for our study.
It was observed with \Su\ twice,
on 2009  for a gross pointing of 109 ks,
and again on 2010 for  99 ks.
These data were briefly analyzed by \citet{EnotoPHD}
and \citet{Enoto10a,Enoto17},
who reconfirmed the  hard X-ray component.
The pulse period was measured at $P=11.00538(5)$ 
as of 2009 August 23 \citep{EnotoPHD, Enoto10b}.

\section{Data Analysis and Results}
\label{sec:ana}

\subsection{Methods}
\label{subsec:ana_methods}

\subsubsection{Selection of the energy range}

In the present work, we analyze the HXD-PIN data
of the three magnetars.
The HXD-PIN instrument utilized cooled Silicon PIN photodiodes,
with its nominal energy band being 10--70 keV \citep{HXD1}.
However, we set the upper energy boundary mostly at $\EU=50$ keV,
because the signal-to-background ratio quickly worsens at higher energies.
We sometimes use $\EU=40$ or 55 keV.
The lower end of the energy band is affected by thermal noise, 
to a varying extent depending on the detector temperature
which in turn is sensitive to the spacecraft attitude 
relative to the Sun \citep{HXD2}.
Since this effect is taken into account by the data reduction pipeline,
the effective lower energy boundary in the archival data is 
typically $\EL= 11- 13$ keV,  depending on the observation.
Based on this information,
and  the actual data behavior,
we choose $\EL$ between 12  and 15 keV.

\subsubsection{Confirmation of the pulsation}
The analysis of each HXD-PIN dataset starts from 
the pulse reconfirmation in an appropriate energy interval.
Specifically, we apply the standard epoch-folding analysis 
to the background-inclusive data,
to produce a periodogram,
over a period range around the pulse period 
that is already confirmed  in the previous works using the same data.
At each trial period, the periodicity significance is evaluated
with the $Z_m^2$ statistics \citep{Z2_83,Z2_93, Makishima23}.
The integer $m$ denotes the maximum harmonic number,
meaning that the Fourier  power of the folded pulse profile 
is summed over the harmonic number $\mu$
running  from $\mu=1$ (fundamental) to $\mu=m$.
At this stage, we set  $m=2$,
because raw pulse profiles of magnetars are often double peaked.
Then, the period $P_0$ that yields the maxim  $Z_2^2$ in the periodogram
is regarded as the pulse period of the object at this observation.

In the present study,
we assume $\dot{P}=0$ for the three sources,
because they  all have 
$\dot P <4 \times 10^{-11}$ s s$^{-1}$ \citep{McGill}.
The expected period change within each observation is at most 8 $\mu$s, 
which is smaller than our period search step (typically 20 $\mu$s).

\subsubsection{Demodulation process}
The essential part of our analysis is
to examine whether the pulsation is subject to PPM  effects.
For this purpose,
we assume that the pulse phase comes and goes sinusoidally,
and to cancel it, 
we modify the arrival time $t$ of each photon (including background) 
to $t-\delta t$, using a correction term
\begin{equation}
\delta t = A \sin(2\pi t/T - \psi_0)~.
\label{eq:demodulation}
\end{equation}
Here, $T$, $A$, and $\psi_0$ are the period, amplitude, 
and initial phase of the assumed PPM, respectively.
Then, we optimize $T$, $A$, and $\psi_0$,
so as to maximize the pulse significance.
This process, called {\it demodulation}, is expected to rectify the pulse phase.
It has been employed successfully in our previous studies (Section \ref{sec:intro}).
We basically set $m=4$, 
because fine structures in the pulse profile will be enhanced.

In practice,  $T$ is varied from 10 ks (or 7 ks) to 100 ks, 
with a step of 0.1 ks to 0.5 ks depending on $T$.
At each $T$, we search for the triplet  $(P, A, \psi_0)$ 
that maximizes $\zz$,
by scanning $P$ with a step of 20 $\mu$s over the uncertainty range of $P_0$,
$A$ from 0 to $\sim 0.15 P$ with a step of 0.1 s,
and $\psi_0$ from $0^\circ$ to $360^\circ$ with a step of $10^\circ$.
If the pulse significance  $\zz$  is enhanced 
significantly at a certain period $T$,
we can claim a PPM detection, 
and identify that $T$ with the slip period $\Tppm$
of Equation~(\ref{eq:slip_period}).

\begin{figure*}
\centerline{
\includegraphics[width=18cm]{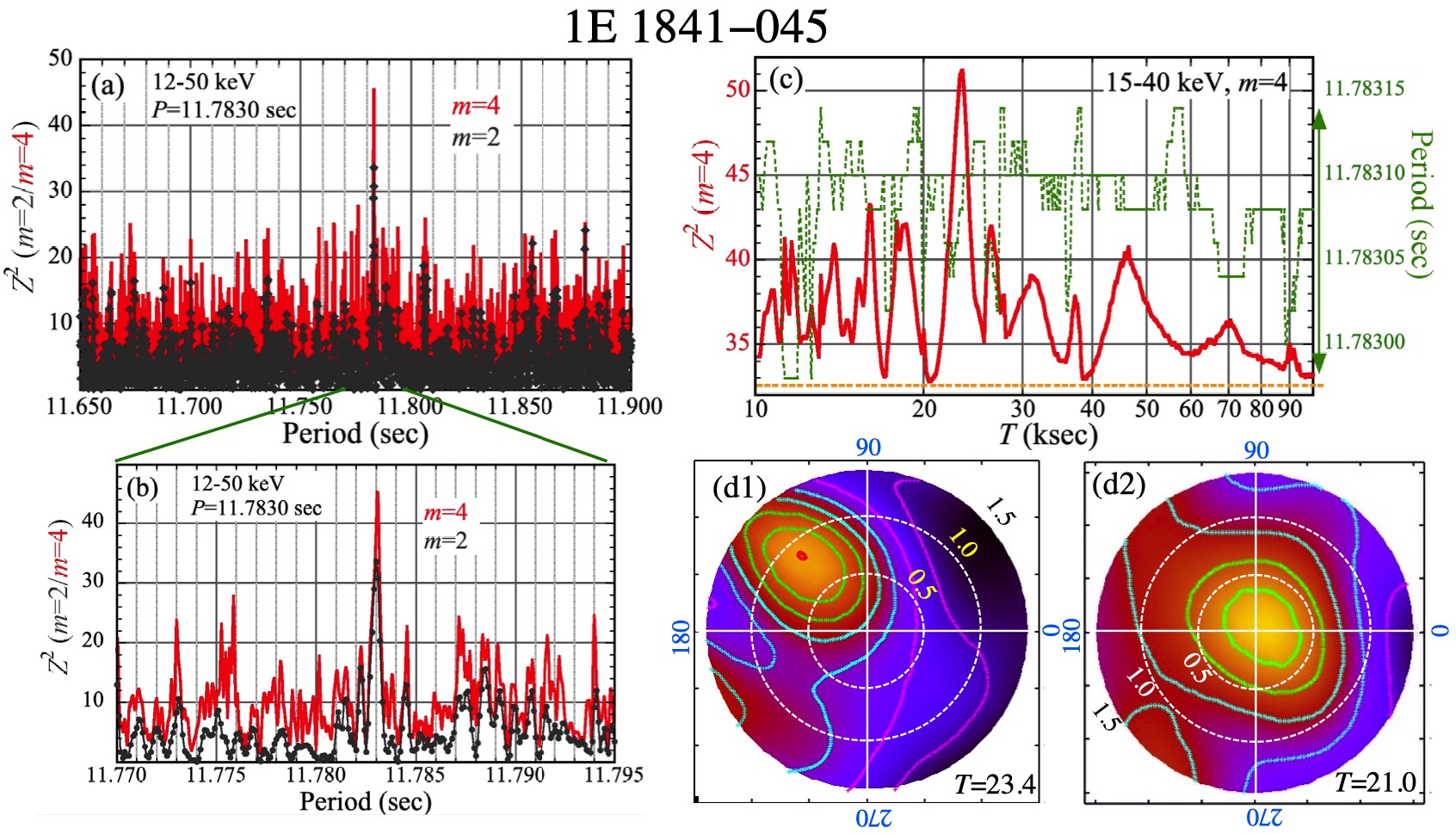}
}
\caption{Results on 1E~1841$-$045.
(a) A 12--50 keV periodogram,
calculated with $m=2$ (black) and $m=4$ (red).
The displayed period interval is  within $\pm 1\%$ of the expected $P_0$.
(b) An expanded view of the periodogram,
within $\pm 0.1\%$ of the detected peak.
(c) A DeMD (red; left ordinate) in 15--40 keV.
The associated values of $P$ are shown in dashed green (right ordinate),
and its scan range  by a vertical green arrow.
The dotted yellow line indicates $\zz$ before the demodulation.
(d1) A color map of $\zz$ on the $(A, \psi_0)$ polar coordinates,
where $P=11.78306$ s and $T=23.4$ ks are both fixed.
Contours are from $\zz=0$ to 50.
(d2) Same as (d1), but using $T=21.0$ ks.
Contours are from 0 to 30.
}
\label{fig:f1_1841}
\end{figure*}

\begin{table*}
\caption{Results of the periodogram and demodulation analyses of the \Su\  HXD-PIN data.$^a$}
\begin{center}
\label{tbl:results}
\begin{tabular}{cccccclccccc}
\hline 
Object (Year)  &Energy $^b$ & $m$& Condition$^c$& $P_0$ (s) & $Z_m^2$ & $T$ (ks)  & $A^d$ & $\psi_0^{e}$  \\
\hline \hline 
\Kes\            &12--50& 2& Raw & 11.78302(11)  & 33.6 & ---  & --- &  --- \\ 
(2006 April)  & 15--40& 4& Demod. &  11.78306 (4)  & 51.2 & $23.4^{+0.5}_{-0.7}$ & 0.8 & $130^\circ$ \\
 \hline 
                    & 12--50& 2& Raw    &   5.76203 (6)  & 17.1& ---  & --- &  ---\\
\SGR\           & 12--50& 4& Demod.&  5.76200 (3)  & 39.7& $16.4^{+0.6}_{-0.5}$ & 0.5 &  $190^\circ$ \\
(2008 Aug.)  & 12--25& 4& Demod.&  5.76199 (3)  & 44.2& $16.4^{+0.6}_{-0.5}$ & 0.6 &  $185^\circ$ \\
                   & 12--55 & 4& Lag    &  5.76199 (3)  & 57.4& $16.6^{+0.4}_{-0.6}$ &0.6 &  $170^\circ$ \\
 \hline 
                 & 13.5--50& 2 & Raw &   11.00527(22)&  49.5   & --- &  --- \\
\RXJ\         & 13.5--28& 4& Demod. & 11.00535(28) & 66.0 &$41.6^{+9.2}_{-9.0}$  &0.7&  $305^\circ$ \\
(2009 Aug.)& 13.5--28& 3+4$^f$ & Demod. &11.00527(15) & 25.7 & $45.7\pm 3.4$   &1.6&  $270^\circ$ \\
                 & 13.5--55&  4 & EDPV &11.00531(17) & 73.8 & $44.2 ^{+2.6}_{-3.1}$  &1.2&  $280^\circ$ \\
 \hline 
                   & 12.5--50&2        & Raw        &11.00594(23) & 45.5&  --- & --- &  --- \\
\RXJ\           & 12.5--28& 4       & Demod.   &11.00595(8)  & 72.5  & $46.5\pm 3.1$ & 1.1 &  $200^\circ$ \\
(2010 Sept.) & 12.5--28&3+4$^f$& Demod.&11.00594(9)   & 31.7  & $46.5\pm 3.1$ &1.3 &  $200^\circ$ \\
                    & 12.5--50  & 4       & EDPV   &11.00596(12)  & 80.3 &  $47.5\pm 3.0$ & 1.1 &  $190^\circ$ \\
\hline 
\hline
\end{tabular}
\begin{footnotesize}
\begin{itemize}
\setlength{\itemsep}{0mm}
\item[$^{a}$] All results assume the period change rate to be 0.
Errors refer to $68\%$ confidence limits.
\item[$^b$] The utilized energy range in keV.
\item[$^c$] ``Raw" and ``Demod." mean without and with the demodulation, respectively.
``Lag" and ``EDPV" mean further corrections with Equation~(\ref{eq:pulse_phase_lag}) and Equation~(\ref{eq:EDPV}), respectively.
\item[$^d$] The modulation amplitude in units of  s, with a typical error by $\pm 0.3$ s.
\item[$^e$] The initial modulation phase, with a typical error by $\pm 20^\circ$.
It should take any value in the $0-360^\circ$ range.
\item[$^f$] The utilized harmonics $\mu$ are given, instead of the maximum harmonic $m$.
\end{itemize}
\end{footnotesize}
\end{center}
\end{table*}

\subsection{\Kes}
\label{subsec:ana_1841}

We first analyzed the data from \Kes,
along the procedure described in Section \ref{subsec:ana_methods}.
As shown in Fig.~\ref{fig:f1_1841}(a) and its expanded view in (b),
a periodogram in 12--50 keV with $m=2$ (black) revealed a clear peak,
at a period of  $P=11.78302(11)~{\rm s}=P_0$.
An $m=4$ periodogram, shown for reference in red, is fully consistent.
This reconfirms the previous pulse detection by \cite{Morii10} from  the same data 
(Section \ref{subsec:obs_1841}).
For some reason, however, the pulsation was not clear 
over the first 12\% of the exposure.
Here and hereafter for this particular dataset,
we hence discard that data portion.

In performing the demodulation analysis, 
we narrowed the energy range from 12--50 keV to  15--40 keV,
so as to focus on the hard X-ray component
referring to the spectral result by \cite{Morii10}.
Moreover, the 15--40 keV range  is generally where HXD-PIN becomes most sensitive;
our first detection of the magnetars' PPM, 
from 4U~0142+61 \citep{Makishima14}, 
was actually attained in this energy interval.
Fig.~\ref{fig:f1_1841}(c) show the obtained result,
where the red curve displays the maximum $Z_4^2$ 
obtained at assumed  $T$;
a plot like this is called a {\em demodulation diagram} (DeMD).
The DeMD reveals a clear peak at $T=23.4^{+0.5}_{-0.7}$ ks,
where  the error (at 68\%) is determined as the points where $\zz$ falls
by 4.72 from the maximum value \citep{Makishima24}.
Since this dataset has about twice longer span than the others,
we extended the DeMD calculation to $T=200$ ks,
but no peak exceeding $\zz \sim 35$ was found.

In  Fig.~\ref{fig:f1_1841}(c), the dashed green line indicates 
the period that maximizes $\zz$ at each $T$,
when $A$ and $\psi$ are also optimized simultaneously.
The DeMD peak  specifies a pulse period of 
$P=11.78306(4)$ s,
which is consistent with $P_0$ from the periodogram.
Panel (d1) shows, as a  polar plot,
the dependence of $\zz$ on $A$ and $\psi_0$,
when $T=23.4$ ks and $P=11.7806$ s are both fixed.
We thus find $A=0.8 \pm 0.3$ s ($6.8 \pm 2.5\%$ of $P_0$),
where the error was estimated by letting $P$ and $T$ float.
These parameters characterizing the DeMD peak
are summarized in Table~\ref{tbl:results},
together with the information from the periodogram.
For reference, when we instead use, e.g.,  $T=21.0$ ks 
where the DeMD shows a valley, 
panel (d2) is obtained;
the maximum $\zz$ occurs at $A\approx 0$,
implying that the pulse phase is not modulated at 21.0 ks.

To evaluate statistical significance of the 23.4 ks DeMD peak,
we conducted a {\em control} study using the actual data,
after \cite{Makishima21b, Makishima24}.
Namely, the same demodulation analysis was repeated,
but scanning $T$ from $0.05$ ks to 5 ks,
assuming that the statistical  fluctuation of $\zz$ is independent of $T$.
To ensure Fourier independence between adjacent samplings,
$T$ was changed with a step  $\Delta T \gtrsim T^2/S$,
where $S=229$ ks is the data span (Table~\ref{tbl:obs}).
We  obtained 3000 trials in $T$, and in only one  out of  them,
$\zz$ exceeded the target value of 51.24, reaching 51.36.
Hence, the probability of finding a peak with $\zz \geq 51.24$
by chance is estimated as $\Pch \sim 1/3000$ for a single trial.
On the other hand, the number of {\em independent} trials in $T$,
contained in Fig.~\ref{fig:f1_1841}(c) 
over the $T=10$ to 100 ks range, 
may be given by the number of independent Fourier waves 
contained therein \citep{Makishima21b},
namely, $S/10 - S/100 \approx 21$.
Then, the chance probability to find a value exceeding $\zz = 51.24$,
at any $T$ between 10 and 100 ks,
is estimated as $1/3000 \times 21 =0.7\%$.
Since the PPM effect is  thus considered real at 99\% confidence,
we can identify  $T=23.4$ ks with $\Tppm$;
\Kes\  becomes a fifth magnetar exhibiting this  phenomenon,
after 4U~0142+61, \oneE, SGR~1900+14, and SGR~1806$-$20.

These results remain mostly unchanged 
when $\EU$ is raised from 40 keV up to $\sim 50$ keV,
beyond which the peak $\zz$ gradually diminishes
due to increasing background contribution.
In contrast, when $\EL$ is lowered to $\lesssim 14$ keV,
the DeMD peak starts to decrease more rapidly,
and gets broader.
As considered in Appendix A,
this is probably because the soft X-ray component begins to contribute to the signal
at energies below 15 keV.
We hence retain the 15--40 keV range.

The 15--40  keV pulse profiles before (black) and after (red) 
the demodulation are shown together  in Fig.~\ref{fig:f1b_1841Pr}(a).
The black profile is rather double-peaked,
wheres the red one exhibits four peaks per cycle,
as often observed among the four preceding magnetars.
Effects of changing $m$ are discussed in Section \ref{subsec:discuss_m},
together with a justification for using $m=4$.
In panel (b) of the same figure,  
the green profile  (also in 15--40 keV) is
derived using about 25\% of the $\Tppm=23.4$ ks cycle,
when the pulse features are  most advanced
according to Equation~(\ref{eq:demodulation}).
Similarly, the orange one uses another 25\% of the $\Tppm$ phase
when the pulse is most delayed.
We discard the remaining 50\% of the data.
The two profiles visualize that the pulse phase 
actually comes and goes  with a period $\Tppm$,
and the demodulation rectifies this behavior.

\begin{figure}
\centerline{
\includegraphics[width=9.cm]{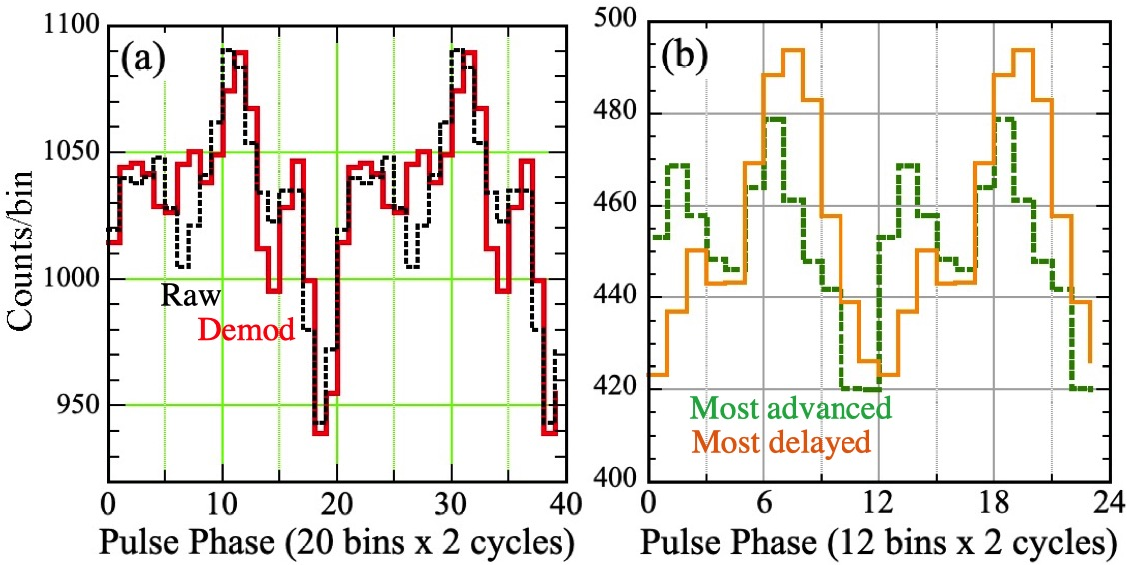}
}
\caption{Pulse profiles of 1E~1841$-$045 in 15--40 keV,
folded at $P=11.78306$ s.
(a) A comparison of 20-bin profiles,
before (black) and after (red) the demodulation.
(b) A comparison of 12-bin profiles before demodulation,
accumulated over two phases of the $\Tppm=23.4$ ks period
(each about 25\% duty), 
when the pulse phase is most advanced (green) and most delayed (orange).
}
\label{fig:f1b_1841Pr}
\end{figure}

\begin{figure*}
\centerline{
\includegraphics[width=16.5cm]{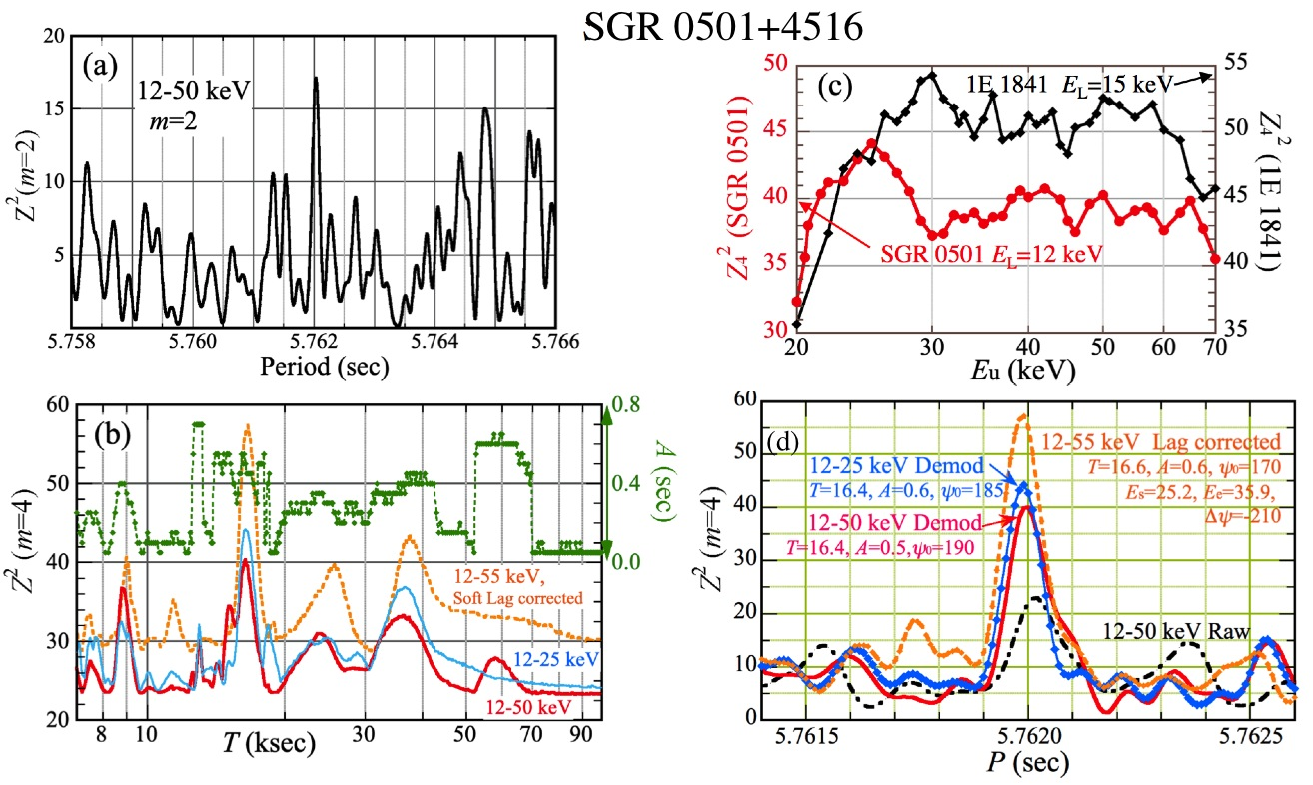}
}
\caption{Results on SGR~0501+4516.
(a) A 12--50 keV periodogram with $m=2$.
The displayed period interval covers  $\pm 0.07\%$ of 
the pulse period reported by \citet{Enoto09}.
(b) DeMDs in 12--50 keV (red) and 12--25 keV (cyan).
The value of $A$ associated with the red DeMD is given in  green (right ordinate).
The dashed orange curve gives $\zz$ by further applying
the soft-lag correction (see text), using 12--55 keV photons.
(c) Dependences of $\zz$ (at $P \approx P_0$ and $T\approx \Tppm$)
on $\EU$ (red) using $\EL=12$ keV,
compared with that of 1E~1842$-$045 (black) with $\EL=15$ keV.
(d) Periodograms ($m=4$) calculated under four different conditions.
Black and red show the 12--50 keV periodograms,
before and after the demodulation, respectively.
Blue is the demodulated 12--25 keV periodogram,
and dotted orange is the demodulated and lag-corrected 12--55 keV periodogram.
The employed parameters  are 
also given ($T$ in ks, $A$ in s, $\psi_0$ and $\Delta \phi$  in degree,
while $\Es$ and $\Ee$  in keV).
}
\label{fig:f2_0501}
\end{figure*}

\subsection{\SGR}
\label{subsec:ana_0501}
For this dataset, we chose $\EL= 12$ keV,
because there were few counts at lower energies.
Fig.~\ref{fig:f2_0501}(a) shows a 12--50 keV periodogram with $m=2$.
It is peaked at a period $P_0=5.76203(6)$ s,
in agreement with that determined by \cite{Enoto09} 
using the same \Su\ data.
In the same way as for \Kes, 
we applied the demodulation analysis to the 12--50 keV data, 
and obtained a DeMD  presented in Fig.~\ref{fig:f2_0501}(b) as a red curve.
It reveals a clear peak with $\zz \approx 40$ at $T\approx 16.4$ ks,
but another peak with $\zz \approx 37$ is seen at 8.8 ks.
The optimum values of $A$, associated with the red DeMD,
are given in green,
which mostly remain at $A\lesssim 0.7$ s, or $\lesssim 1.2 P_0$,
as in the preceding four magnetars.

In Fig.~\ref{fig:f2_0501}(b), the cyan DeMD was 
derived with the energy range limited  to 12--25 keV.
The $T\approx 16.4$ ks peak became more outstanding,
and the 8.8 keV sub-peak diminished.
We can now identify $T \approx 16.4$ ks with $\Tppm$,
but the difference between the red and cyan DeMDs must be explained.
We hence fixed $\EL=12$ keV, and examined how the DeMD peak height depends on $\EU$.
Then, from $\EU=20$ keV up to $\EU \approx 25$ keV,
the peak became higher  as plotted in Fig.~\ref{fig:f2_0501}(c) in red,
obviously due to the increased photon number.
Beyond $\EU \approx 25$ keV,
$\zz$ decreases rapidly till $\approx 30$ keV,
where it  hits a floor, or even recovers slightly.
This behavior on one hand agrees with the relation between the red and cyan DeMDs,
but is puzzling on the other hand,
because $\zz$ would decrease {\em monotonically} beyond a certain energy
above which the  background dominates.
In fact,  $\zz$ of \Kes, shown for comparison in black,
behaves  roughly in that way.
We suspect that  the pulsation of \SGR\  is affected, 
from $\sim 25$ keV up to $\sim 30$ keV,
by some timing disturbance that is added on top of the PPM.

This issue was studied in Appendix B,
where we found that the pulse phase of \SGR\ shifts with energy, 
from $\approx 25.2$  keV up to  $\approx 35.9$ keV,
almost by half a  cycle.
Although it remains ambiguous whether this behavior  
should be modeled as ``soft lag'' or ``hard lag",
we adopted the  soft-lag option  (Appendix B),
and corrected the arrival times of individual photons
(including background)  for this lag effect
using Equation~(\ref{eq:pulse_phase_lag}).
This timing correction operates as  a function of the pulse phase,
whereas  Equation~(\ref{eq:demodulation}) works
as a function of the modulation phase.
As a result, the peak at $T\approx 16.4~{\rm ks}=\Tppm$ 
in the 12--50 keV DeMD became higher
from $\zz \approx 40.0$ to $\zz \approx 53.6$.
As  shown  by an orange curve in Fig.~\ref{fig:f2_0501}(b),
it further increased to $\zz \approx 57.4$,
when $\EU$ is raised to 55 keV.

Since the pulse-phase lag effect must be  independent of the modulation phase,
it should be  observed to some extent before the demodulation.
A  comparison of raw pulse profiles in different energy ranges in fact 
gave some evidence for the same effect, but was not convincing enough,
probably because the pulsation is rather weak before the demodulation.
Hence, the pulse-phase lag is  considered to be a second-order  perturbation,
and becomes noticeable after correcting the data for the PPM 
which is regarded as the first-order effect.

\begin{figure}
\centerline{
\includegraphics[width=6.3cm]{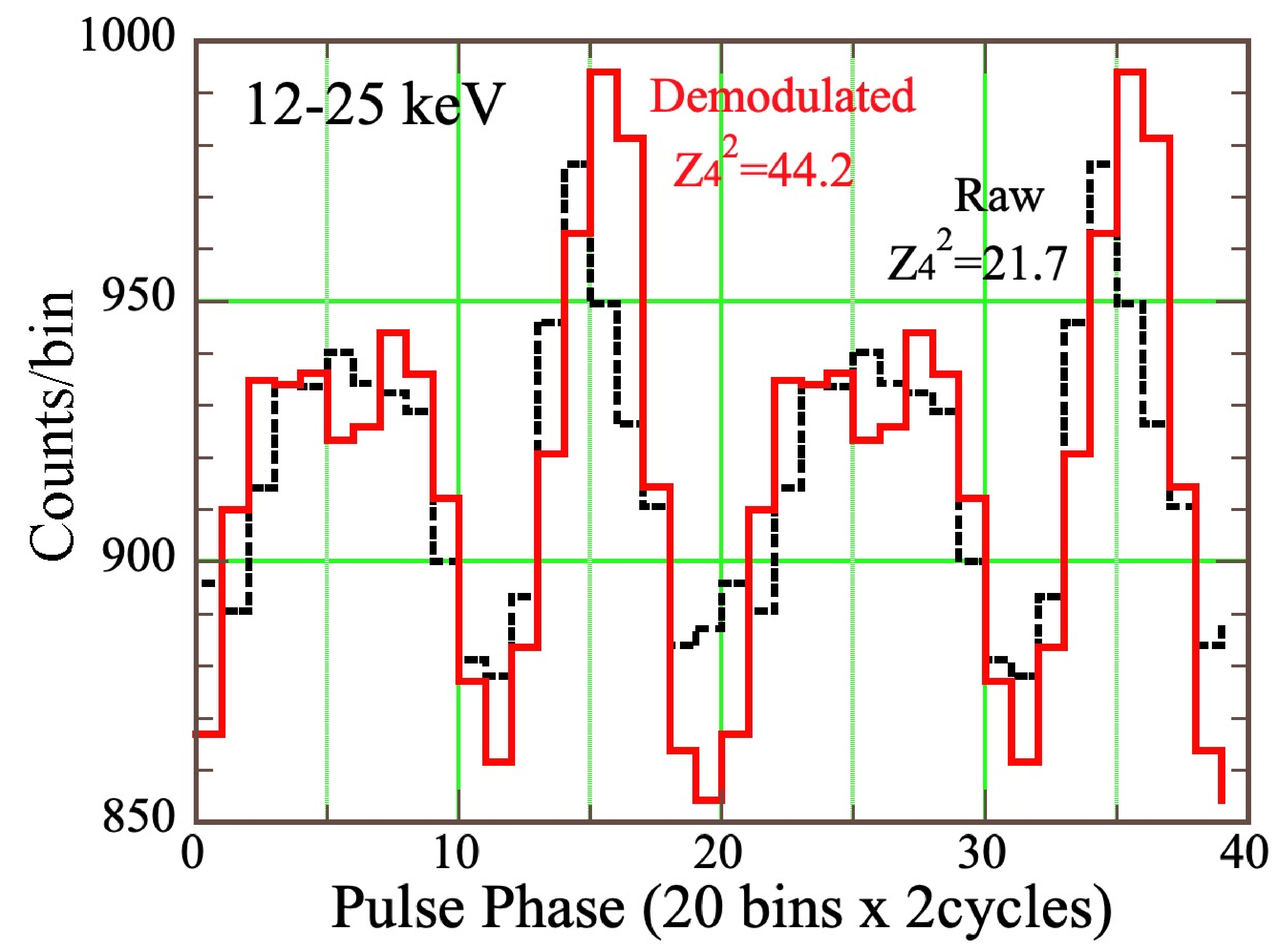}
}
\caption{
Folded pulse profiles of \SGR\ in 12--25 keV,
before (black) and after (red) the demodulation.
}
\label{fig:f2b_0501Pr}
\end{figure}

While  $\zz$ increased significantly through 
the two-stage timing corrections (for the PPM and the lag effect),        
the best pulse period $P_0$ has remained nearly the same.
This is clear in Fig.~\ref{fig:f2_0501}(d),
which compares periodograms computed under four different conditions (see caption).
The peak heights of the red, blue, and orange periodograms
are identical to the DeMD peak heights of the same color
in panel (b).

\begin{figure*}
\centerline{
\includegraphics[width=17.5cm]{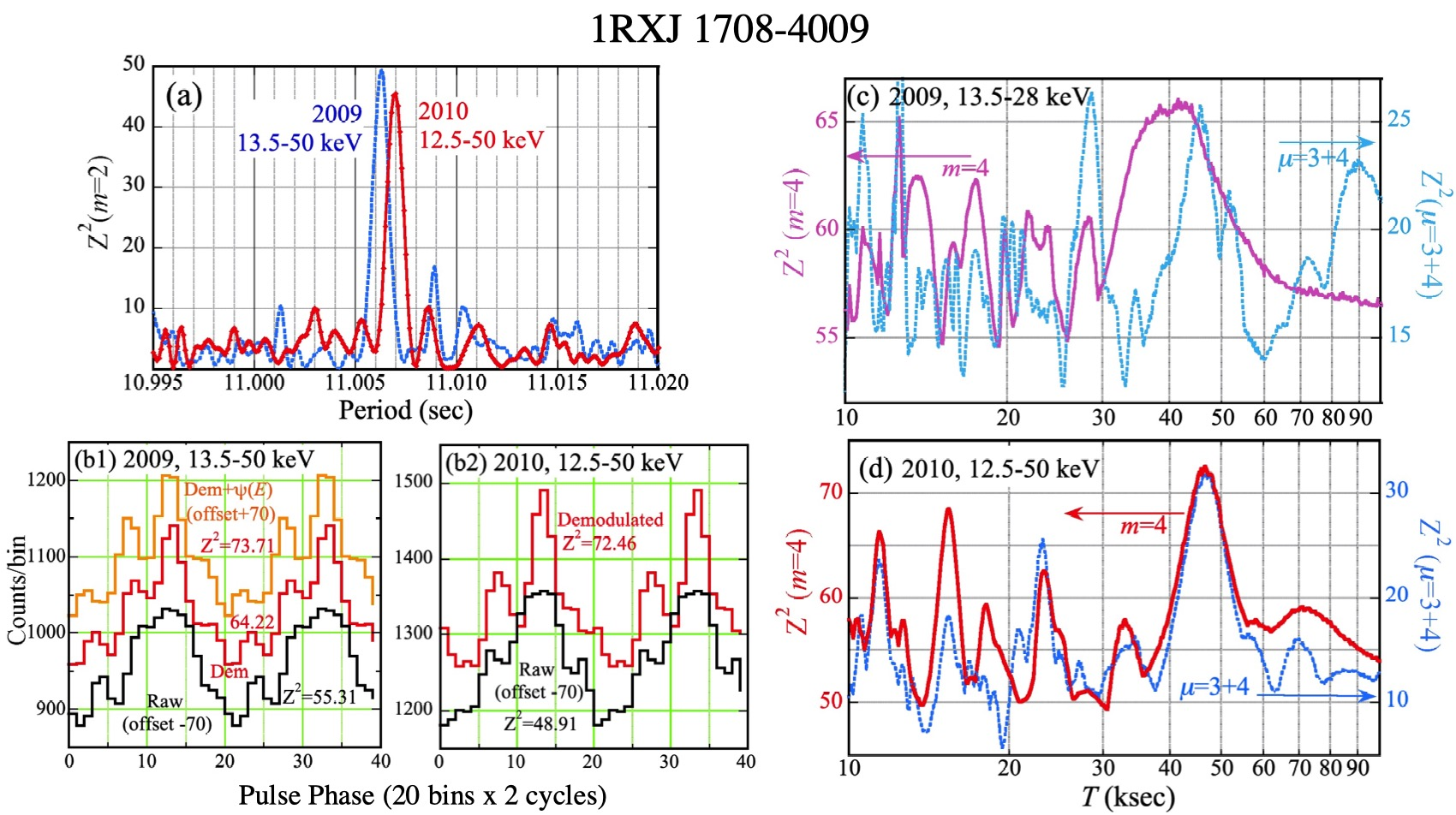}
}
\caption{Results from the two datasets of  \RXJ.
(a) Periodograms with $m=2$  in 2009 (blue; 13.5--50 keV)
and 2010 (red;12.5--50 keV).
(b1) 13.5--50 keV pulse profiles in 2009,
before (black) and after (red)  the demodulation.
The orange one is further corrected for the EDPV disturbance,
and is given a vertical offset by 70.
(b2) Same, but from the 2010 data in 12.5--50 keV.
(c) DeMDs in 13.5--28 keV from the 2009 data,
using $m=4$ (purple; left ordinate) 
and $\mu=3$ plus $\mu=4$ (cyan; right ordinate).
(d) DeMDs from the 2010 data.
The red curve is for $m=4$, 
while  the cyan one is for $\mu=3+4$,
both in 12.5--50 keV.
}
\label{fig:f3_1708}
\end{figure*}

The significance of the 16.4 ks PPM 
was estimated via the same control study as for \Kes,
in 12--25 keV,  the widest energy range free from the phase-lag disturbance.
We obtained  $\Pch < 6.3 \times 10^{-4}$ for a single trial,  
as $\zz$  from the total 1600 trials were all  $\leq 43.02$,
and did not reach  $\zz=44.17$.
Since the effective number of trials in $T$  from 7 ks to 100 ks 
is estimated as  $\approx 15$  (for $S=110$ ks),
the  post-trial chance probability becomes 
$\Pch < 6.3 \times 10^{-4} \times 15 \approx 1\%$.
Thus, the PPM at $\Tppm \approx 16.4$ ks is 
statistically significant at 99\% confidence, as in \Kes.
It is more difficult to quantify the significance of the phase-lag effect,
but we regard it as real,
because essentially the same effect has  been  observed from two magnetars,
\oneE\  \citep{Makishima21a} and SGR~1900+14  \citep{Makishima21b},
as well as  the gamma-ray binary LS~5039  \citep{Makishima23}.
Discussion continues in Section \ref{subsec:discuss_perturbations}.

Fig.~\ref{fig:f2b_0501Pr} presents the pulse profiles
of \SGR\ before (black) and after (red) the demodulation,
using  the 12--25 keV range 
which is free from the  phase lag effects.
The raw profile (black)  is  double-peaked,
whereas three peaks have developed in the red profile.
These results are consistent with a finding
to be described  in Section \ref{subsec:discuss_m},
that the $\mu=2$ and  $\mu=3$ components dominate
the demodulate profile of this  object,
with weak contributions from $\mu=1$ and $\mu=4$.
As  given in Appendix B (Fig.~\ref{fig:f_AppB_Pr}), 
the demodulated pulse profile  in a wider 12--55 keV range is very similar.

In this way, we have detected a statistically significant PPM effect
with a slip period of $\Tppm \approx 16.4$ ks from \SGR,
which is now the sixth magnetar showing this behavior.
Furthermore, a phase-lag phenomenon has been noticed 
as a second-order timing perturbation,
operating in the 25.2 to 35.9 keV interval.

\subsection{1RXS J1708$-$4009}
\label{subsec:ana_1708}

The two datasets of \RXJ\ were acquired,
in 2009 and 2010,
under similar exposures ($\sim 100$ ks in gross) 
and similar source intensities.
They both have  $\EL \approx 12.5$ keV,
but the pulse period in 2009 below $\approx 13.5$ keV
is inconsistent with  that  above 13.5 keV,
presumably due to enhanced thermal noise.
(Such effects were not seen in the 2010 data.)
Therefore, we set $\EL = 13.5$ keV and $\EL =12.5$ keV, 
for the 2009 and 2010 data, respectively.

The results from these data  are summarized in Fig.~\ref{fig:f3_1708},
where panel (a) shows $m=2$ periodograms on the two occasions.
The pulsation is detected in both periodograms
with similar peak heights and widths.
The period in 2009 (Table~\ref{tbl:results}) is consistent with
that in  \cite{EnotoPHD} who analyzed the same data.
The period difference by $\Delta P \approx 0.7$ ms between the two periodograms
implies $\dot P \sim 2 \times 10^{-11}$ s s$^{-1}$,
in agreement with the literature \citep{McGill}.
Raw  pulse profiles in 2009 and 2010 are 
given by black histograms in panels (b1) and (b2), respectively.
They look alike, with a broad single peak
with a few minor substructures.
Panels (c) and (d) give DeMDs from the 2009 and 2010 data, respectively.
Below, we first describe the 2010 results,
which are much simpler than those in 2009.

\subsubsection{Demodulation of the 2010 data}
The red curve in Fig.~\ref{fig:f3_1708}(d)
shows the $m=4$ DeMD in 2010,
using the 12.5--50 keV interval. 
A prominent peak  emerged at $T \approx 46$ ks,
together with several weaker peaks at $T < 30$ ks.
Another DeMD in blue sums the  $\mu=3$ and  $\mu=4$  
powers of the profile, instead of using $m=4$
(i.e., $\mu=1$ to $4$ summed up).
Except a systematic difference in $\zz$ by $ \sim 40$
which comes from the $\mu=1$ and $\mu=2$ powers,
the two DeMDs behave  similarly, 
including the  $T \approx 46$ ks peak
to be identified with $\Tppm$.
In this dataset, the demodulation process is thus seen
to selectively enhance  the 3rd and 4th harmonic powers of the profile.
This inference reconfirms  Fig.~\ref{fig:f3_1708}(b2),
where a 4-peak feature develops through the demodulation.

The long $\Tppm$ of \RXJ\ allows us to visualize the PPM effect.
In Fig.~\ref{fig:RXJ1708_10_timelapse}, 
we sorted 12.5--50 keV photons in the 2010 data,
as a function of the pulse phase and elapsed time.
The derived photon counts are color coded,
after re-normalizing the ten rows so as to have 
the same average and the same variance.
Through the total time span of 99 ks,
the pulse ridge, seen in yellow, clearly wiggles about twice,
with an amplitude of $\sim \pm 0.1$  pulse cycles.
These reconfirm the DeMD results of $\Tppm \approx  46$ ks
and $A =1.2$ s $\approx 0.1 P$.

\begin{figure}
\centerline{
\includegraphics[width=6.3cm]{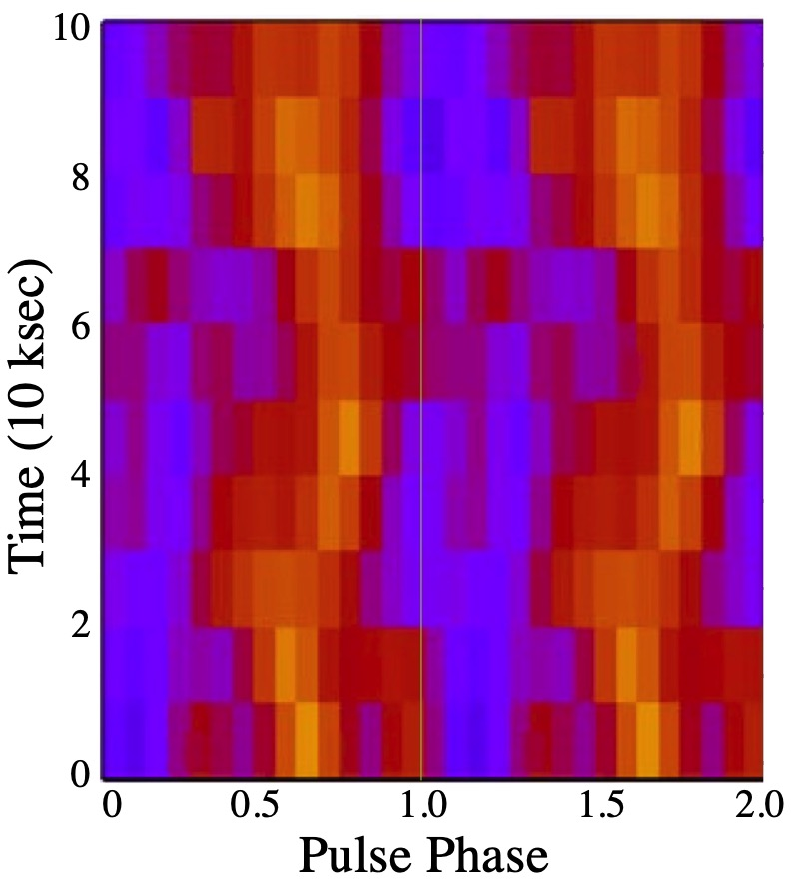}
}
\caption{A color intensity map of 12.5--50 keV photons of \RXJ\ in 2010,
displayed as a function of the pulse phase (abscissa) and the elapsed time (ordinate).
}
\label{fig:RXJ1708_10_timelapse}
\end{figure}

In a similar way to the preceding two sources,
we evaluated the statistical significance of the 46 ks DeMD peak  in 2010,
and obtained $\Pch \sim 1.3\%$ considering the samplings in $T$.
Therefore, this AXP can be
regarded as the seventh magnetar
that exhibits the PPM phenomenon,
although our final conclusion must await the 2009 results.

\subsubsection{Demodulation of the 2009 data}
\label{subsubsec:ana_1708_2009}

As a first-cut trial on the 2009 data, we calculated a 13.5--50 keV DeMD.
As $T$ varied from 10 ks to 100 ks,
$\zz$ fluctuated between $57$ and $71$,
but no clear peak appeared.
The amplitude remained $A<0.5$ s,
contrary to the 2010 DeMD 
in which  $A \approx 1.2$ s is recorded at   $T \sim 46$ ks.
Thus, the pulsation in the 2009 data is significant
already before the demodulation, 
and the data do not prefer particular values of $T$.
This result, apparently  inconsistent with that in 2010,
may be interpreted in several alternative ways.
As one possibility, the 46 ks peak in the 2010 DeMD 
was a transient feature,
and was not present in 2009.
Or else, the 46 ks PPM resides in the 2009 data, too,
but undetectable due to a small amplitude,
as in the 2014 \NuS\ observation of 4U~0142+61 \citep{Makishima19}
wherein its 55 ks PPM had  such a small amplitude as $A \approx 0.02 P$
that the effect was marginally detectable even with \NuS.
Yet a third possibility is that the 46 ks PPM,
though present  with $A \sim 1$ s, 
is affected in the 2009 data  by some additional timing disturbance.

To examine the third option above,
we first suspected the pulse-phase lag  as  in \SGR,
but no evidence was found in the data.
In search for alternative clues, we produced  DeMDs
by changing the energy band and $m$.
Then, in a 13.5--28 keV DeMD with $m=4$,
a broad hump emerged over $T=33-50$ ks 
as shown in magenta  in Fig.~\ref{fig:f3_1708}(c),
(This hump diminished for $\EU > 28$ keV).
When we retain the 13.5--28 keV range
but instead use $\mu=3+4$,
like the blue curve in panel (d),
a DeMD shown in cyan in panel (c) was derived;
it resembles the 2010 results,
except differences in the  vertical scales
and the presence of another  peak at $T \approx 29$ ks.
The magenta hump and the cyan peak demand 
$A \approx 0.5$ s and $A \approx 1.5$ s, respectively.
About $2/3$ of the cyan peak at $T \approx 46$ ks
comes from $\mu=3$, and about 1/3 from $\mu=4$.
Thus,  the PPM with $\Tppm \approx 46$ ks  is likely to
lurk in the 2009 data with an amplitude that is not too small,
but the pulsation may suffer yet another type of
second-order  timing perturbation
that depends on the energy and the harmonic number.

\begin{figure}
\centerline{
\includegraphics[width=8.4cm]{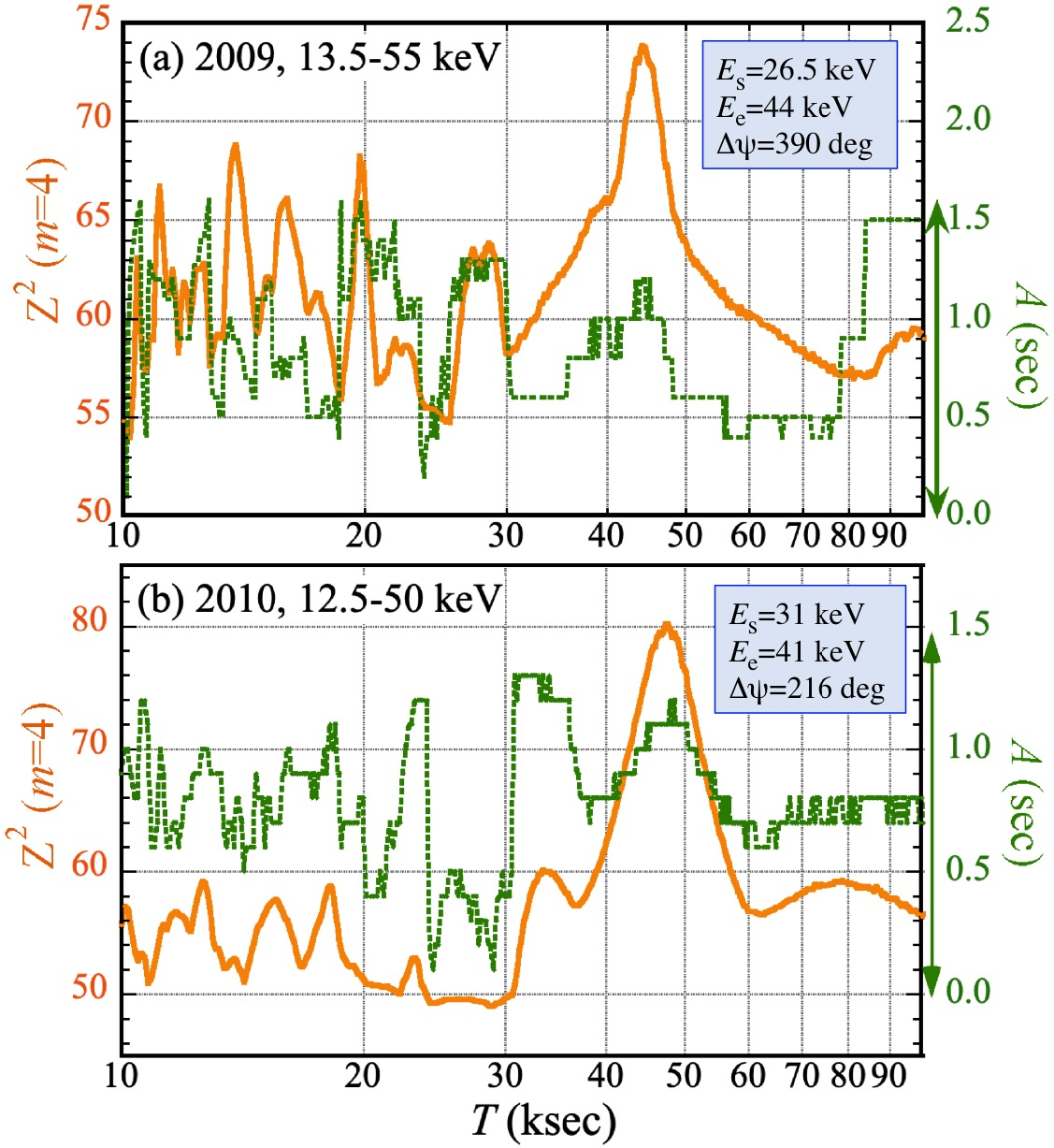}
}
\caption{DeMDs (orange) of \RXJ,
obtained by the demodulation and  the EDPV correction (Appendix C).
Panel (a) is from the 13.5--55 keV data in 2009, 
and (b)  from the 12.5--50 keV data in 2010.
The associated values of $A$ are shown in green.
The employed EDPV parameters are given in the figures.
}
\label{fig:f4_1708_EDPV}
\end{figure}

\subsubsection{Corrections for the energy dependence in $\psi_0$}
\label{subsubsec:ana_1708_EDPV_correction}

The above issue  with the 2009 data was further examined in Appendix C.
There,  we found
that the modulation phase $\psi_0$ in Equation~(\ref{eq:demodulation}) 
changes with energy, from $\approx 26.5$ keV to $\approx 44$ keV,
by $\Delta \psi =390^\circ \pm 50^\circ$.
This is  essentially the same phenomenon as observed  from
\oneE\  and SGR~1900+14 (see Appendix C);
it was then named ``Energy Dependent Phase Variation (EDPV)".
Employing Equation (\ref{eq:EDPV}) in Appendix C,
we applied an energy-dependent timing correction to $\psi_0$, 
or ``EDPV correction", for the 13.5--55 data in 2009, 
and obtained a DeMD shown in  Fig.~\ref{fig:f4_1708_EDPV}(a).
It exhibits a very strong peak reaching $\zz=73.8$,
at  $T=44.4 \pm 2.8$ ks which is consistent with  that in 2010  (Table~\ref{tbl:results}).
Furthermore, the peak is no  longer very broad.
Even though the EDPV parameters might still be somewhat ambiguous,
the 2009 data can thus  be brought into a consistency with those in 2010,
via the EDPV correction which has been applied successfully 
to the \NuS\ and \Su\ HXD data of the two other magnetars.
We hence reconfirm that \RXJ\ shows the PPM with $\Tppm \approx 46$ ks.

An immediate question is whether the EDPV effect 
also lurks in the 2010 data or not.
The answer is yes;  via the same correction, 
but employing  $E_{\rm s} = 31$ keV, $E_{\rm e} = 41$ keV, 
and $\Delta \psi=216^\circ$, 
the 12.5--50 keV DeMD becomes as presented in  Fig.~\ref{fig:f4_1708_EDPV}(b).
The main peak increased to $\zz \approx 80.3$ (from 72.5),
and the secondary peaks  at $T<30$ ks became much suppressed.
The amplitude  $A \approx 1.0$ s is similar between 2009 and 2010,
as given by the green line in Fig.~\ref{fig:f4_1708_EDPV}.
The EDPV perturbation is thus present in the 2010 data as well,
but it affects a  narrower energy range (31--41 keV)
than in 2009 (26.5--44 keV),
with $\Delta \psi \approx 216^\circ$  which is smaller than in 2009
($\Delta \psi \approx 390^\circ$).
These may explain the reason why the demodulation of 
the 2010 data  (Fig.~\ref{fig:f3_1708}d)
worked even without the EDPV correction.

As to the value of $\Tppm$,
we refrain from using Fig.~\ref{fig:f4_1708_EDPV},
because of possible residual ambiguities in the EDPV parameters (in both datasets).
Instead, we adopt the red DeMD in Fig.~\ref{fig:f3_1708}(d)
from the 2010 data, and quote $\Tppm =46.5 \pm 3.1$ ks (Table~\ref{tbl:results}). 

\section{Discussion}
\label{sec:discuss}
\subsection{Summary of the data analysis}
\label{subsec:discuss_summary}

We studied four  \Su\ HXD observations of three magnetars;
one for \Kes, another for  \SGR, and the other two for  \RXJ.
The source pulsation in each observation was confirmed
at a period that agrees with previous measurements
(Table~\ref{tbl:results}).

Subsequently, we applied the demodulation analysis to each dataset,
and successfully detected the PPM (pulse-phase modulation) from all three objects,
at a  slip period of
$\Tppm=23.4^{+0.5}_{-0.7}$ ks from \Kes, 
$16.4^{+0.6}_{-0.5}$ ks from \SGR, 
and $46.5\pm 3.1$ ks from \RXJ\ 
(Table~\ref{tbl:results}).
These  PPM effects are considered real,
because they all have  $\Pch \sim 1\%$,
when considering the trials in $T$, $A$, and $\psi$, 
as well as in $P$ over the uncertainty range of $P_0$.
Including the previously confirmed four magnetars,
now we have a list of seven magnetars
that exhibit the PPM phenomenon.

The three magnetars studied here share 
the following PPM properties in common;
these are also seen in the preceding four objects.
(i) The modulation amplitude is  $A = (0.06-0.16) P$.
(ii) $\Tppm$ is determined with $\approx 3\%$ accuracy. 
(iii) The demodulated pulse profiles have 3 to 4 peaks,
and  are not strongly energy dependent,
at least over the $\approx 12$ to $\approx 50$ keV range.
(iv) The $P_0/\Tppm$ ratio is in  a narrow range, $(2.4-5.0)\times 10^{-4}$.

Yet another important property of the previous four magnetars 
is the exclusive association of their PPM with the spectral hard component.
Although the lack of information below 10 keV in the present work
hampered us to confirm this property in \SGR\ and \RXJ,
we obtained in Appendix A some indication from \Kes,
that its PPM disappears in the 10--15 keV interval,
where the soft-component contribution may not be negligible.
Therefore, we presume that the PPM effects
detected from the present three objects
are also associated with their hard X-ray component.



While the demodulation process certainly strengthened 
the pulse significance in the three objects,
we had often to incorporate corrections to
additional timing perturbations.
These findings are discussed in Section \ref{subsec:discuss_perturbations},
in comparison with past observations of the same effect.

\begin{figure}
\centerline{
\includegraphics[width=5cm]{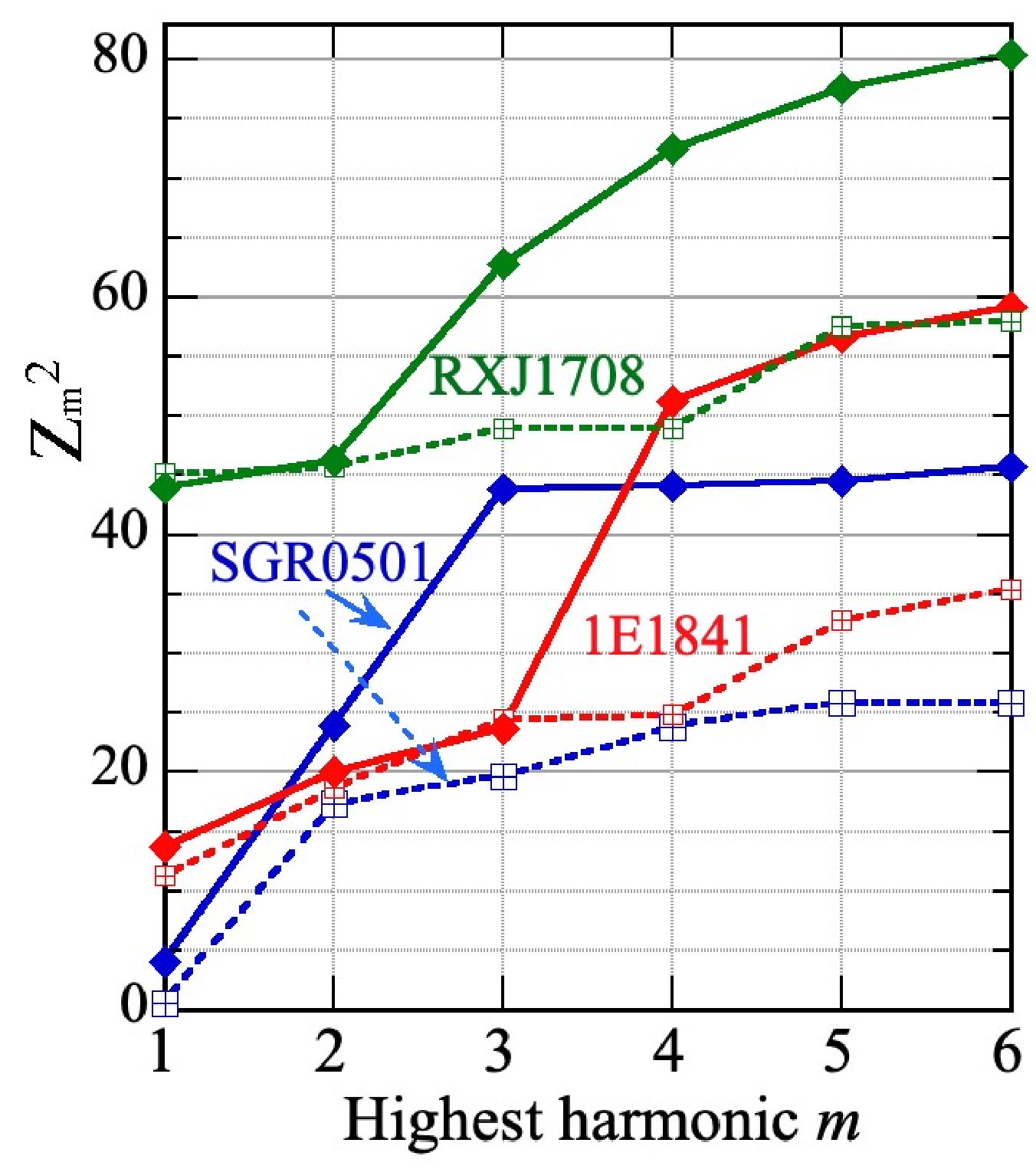}
}
\caption{Dependences of $Z_m^2$ (the maximum value at $P \approx P_0$)
on the maximum harmonic number $m$.
Red, blue, and green indicate \Kes\ (15--40 keV),
\SGR\ (12--25 keV), and \RXJ\ in 2010 (12.5--50 keV), respectively.
The results before the demodulation are shown by dotted lines,
whereas those after by solid lines.
}
\label{fig:ProfileHarmonics}
\end{figure}

\begin{figure}
\centerline{
\includegraphics[width=8cm]{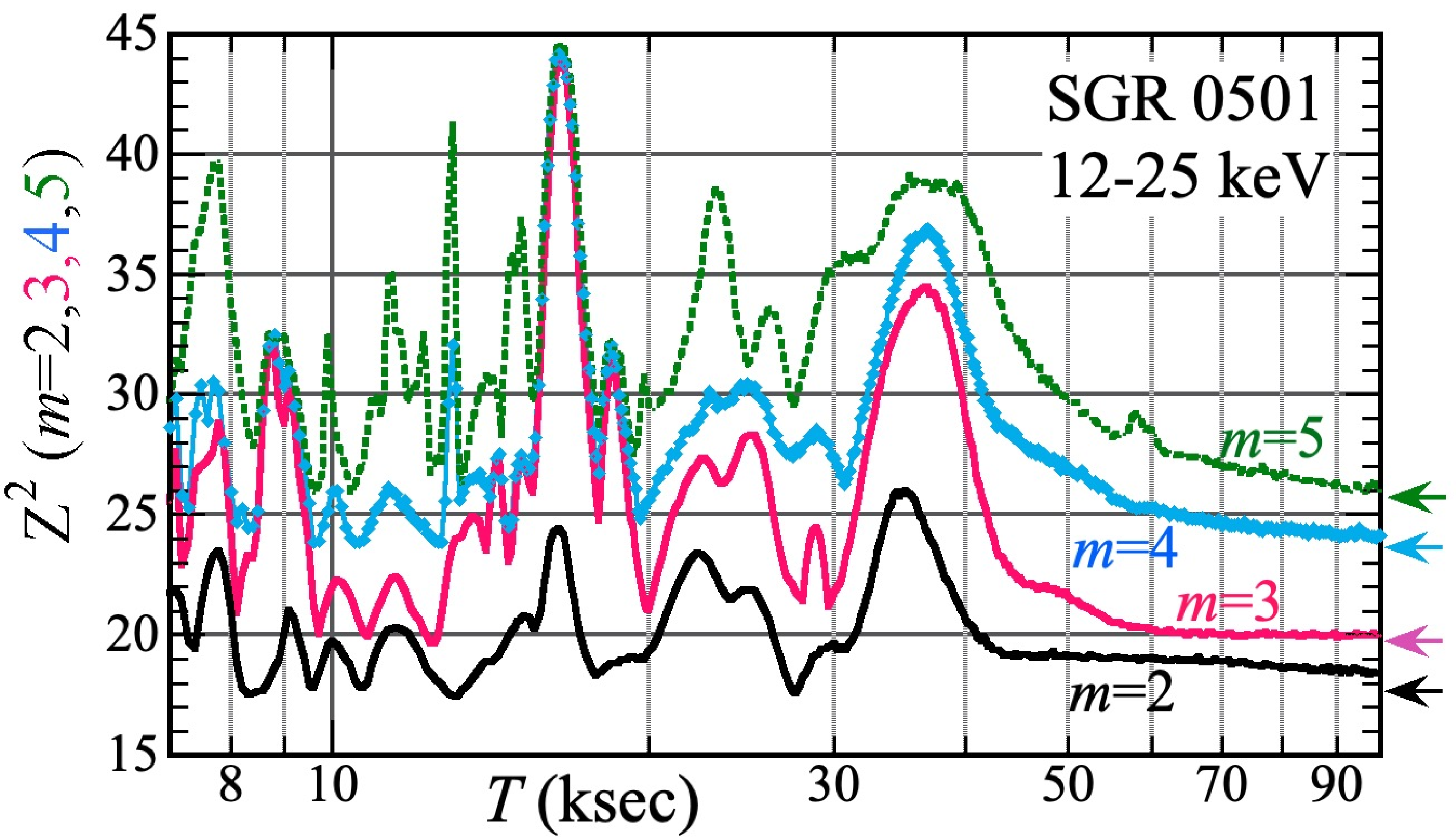}
}
\caption{DeMDs of \SGR\ in 12--25 keV,
calculated with $m=2$ (black), $3$ (magenta), $4$ (cyan), and $5$ (green).
The $m=4$ result  is identical to the cyan curve in Fig.~\ref{fig:f2_0501}(b).
The arrows near the right ordinate indicate the $\zz$ values of the pulsation
 before the  demodulation.
}
\label{fig:SGR0501_Tscan_multi-m}
\end{figure}

\subsection{Contributions  from  various  harmonics }
\label{subsec:discuss_m}
To justify our choice of $m=4$,
and examine how various harmonics of the pulse profile 
contribute to the overall pulse significance,
we computed $Z_m^2$ of the three sources at 
$P\approx P_0$ and $T \approx \Tppm$, 
by changing $m$ from 1 to 6.
Fig.~\ref{fig:ProfileHarmonics} shows the results,
with and without the demodulation.
The energy ranges, given in the caption, were  selected  
so as to avoid the two types of timing perturbations.
For the same reason, the 2009 data of \RXJ\  were not used here.
In maximizing the demodulated $Z_m^2$,
for each $m$ we readjusted $P, T, A$, and $\psi$,
within their appropriate error ranges.
The three sources  behave rather distinctly
in Fig.~\ref{fig:ProfileHarmonics}.

The result on \Kes\  before the demodulation is dominated by the $\mu=1$ component.
The demodulation makes the $\mu=4$ power overwhelming,
but little affects the other ($\mu \ne 4$) components.
This is consistent with the pulse profiles presented in Fig.~\ref{fig:f1b_1841Pr}(a).

In  \SGR, the $\mu=2$ power is dominant before the demodulation,
whereas $\mu=3$ becomes similarly strong afterwards.
This is  already expected from the profiles in Fig.~\ref{fig:f2b_0501Pr}.
Interestingly, unlike the case of \Kes,
the $\mu=4$ power remains weak throughout.
This property is more clearly seen in Fig.~\ref{fig:SGR0501_Tscan_multi-m},
where DeMDs of \SGR\ with different $m$ are superposed together, where 
the DeMDs  with $m=3$, $m=4$ and $m=5$
 exhibit nearly the same peak height.
If we use $m=3$ instead of $m=4$,
$\Pch$  decreases from $1.3\%$ to $0.2\%$,
due to the reduced degree of freedom.

In the 2010 data of \RXJ,
the fundamental ($\mu=1$) contribution is dominant at first.
The demodulation process selectively enhances
the $\mu=3$ and $\mu=4$ powers to a similar extent,
but not the other components.
Again, these properties agree with the behavior
of the pulse profiles in Fig.~\ref{fig:f3_1708}(b2).
Since the pulse profiles in 2009 are not much different,
these harmonic contributions found in 2010 
are considered to apply also to the 2009 data.

As seen so far, the $m$-dependence of the demodulated pulse profile
differs considerably among the three sources.
Nevertheless,  the following properties are common to them.
(i) Neither  $\mu=1$ nor $\mu=2$ powers are 
strongly affected by the PPM.
(ii) The demodulation significantly enhances
the $\mu=3$ or $\mu=4$ power, or both.
(iii) Fourier components with $\mu \ge 5$ are  weak,
both before and after the demodulation.

These outcomes from  Fig.~\ref{fig:ProfileHarmonics} 
justify our use of $m=4$, although $m=3$ can be even better 
sometimes (for \SGR\ in the present case).
In addition, the properties of the $\mu \le 2$ and $\mu \ge 3$ powers
are too distinct to be ascribed to a simple expectation,
that components with higher $\mu$ would be
more strongly smeared by the PPM.
As a possible explanation,
the hard-component emission may consist of two components;
one has a relatively broad beam pattern with little asymmetry around $\hat x_3$,
producing  the $\mu=1$ and 2 powers,
whereas the other has three to four collimated beams
(responsible for the $\mu=3$ and 4 components)
that are tilted off $\hat x_3$.
As already discussed briefly in \citet{Makishima21a},
these results will  provide a valuable insight into 
the emission mechanism of the hard X-ray component of magnetars,
but further examination of this subject is beyond the scope of the present paper.

\begin{table*}
\caption{Summary of PPM measurements and estimates of $\Bt$.}
\begin{center}
\label{tbl:PPM_summary}
\begin{tabular}{lccccccccc}
\hline 
Source  & Year  & $P_0$  & $\Tppm$ & $P_0/\Tppm$  &${B_{\rm t}'}^a$& ${B_{\rm d}}^b$
      & $B_{\rm t}'/B_{\rm d}$ &${\tau_{\rm c}}^c$ & Ref.$^d$\\
    &          &  (s) &     (ks)  & $(10^{-4})$  & $(10^{16}$ G) & $(10^{14}$ G) & & (kyr)\\
\hline \hline 
4U 0142+61 & 2009  & 8.689& $55 \pm 4 $ &1.6    &  1.3 &1.3 & 100    &68 & [1]\\                             
\SGR        & 2008 &  5.762 & $16.4^{+0.6}_{-0.5}$ & 3.5 & 1.9& 1.9  &100 & 15 & [2]\\
\RXJ         & 2010 &  11.006 &$46.5\pm 3.1$ &2.4  & 1.5 &  4.7 &32 &  9.0 & [2]\\
\Kes              & 2006  & 11.783 & $23.4^{+0.5}_{-0.7}$ & 5.0 &2.2 &  7.0  &31 & 4.6 & [2]\\
SGR 1900+14 & 2016 & 5.227  & $40.5\pm 0.8 $ & 1.3  & 1.1 & 7.0 &16 & 0.90 &[1] \\
\oneE               &2016 &2.087 & $36.0 \pm 2.3$ &0.58 & 0.76 &3.2 & 24 &0.69 & [1]   \\
SGR 1806-20 & 1993 & 7.469 & $16.435 \pm 0.024$ & 4.5 &2.1&20 &11 &0.24 & [1] \\
\hline 
\end{tabular}
\end{center}
\begin{footnotesize}
\begin{itemize}
\setlength{\itemsep}{0mm}
\item[$^a$] Defined in  Equation~(\ref{eq:Bt'}).
\item[$^b$] The surface dipole magnetic field strength,  taken from \cite{McGill}.
\item[$^c$] The characteristic age $\tau_{\rm c} \equiv P/2 \dot {P}$ \citep{McGill}. 
\item[$^d$] References: [1] \citet{Makishima24}, [2] This work.
\end{itemize}
\end{footnotesize}
\end{table*}

\subsection{Evolutions of the dipole and toroidal fields}
\label{subsec:BtBd_evolution}

\subsubsection{Calculation of $\Bt$}

We  now have  the PPM phenomenon detected from seven magnetars,
as summarized in Table~\ref{tbl:PPM_summary}
which updates table 3 of  \cite{Makishima24}.
The seven objects comprise all the three major magnetar subclasses;
two prototypical SGRs (SGR~1806$-$20 and SGR~1900+14),
three representative AXPs (4U~0142+61, \Kes, and \RXJ),
and two transient magnetars (\oneE\ and \SGR).
Thus,  the PPM is  inferred to be a common property of magnetars.
In addition, the basic PPM properties, such as
items (i) through (iv) in Section \ref{subsec:discuss_summary},
are mostly common to all these sources.
Therefore, we are likely to be observing a single physical phenomenon,
namely, the free precession of axially defamed NSs.
The  following discussion assumes axial symmetry for simplicity,
although  possible triaxial deformation \citep{Gao23}
cannot be excluded.

Eliminating $\epsilon$ from  Equations~(\ref{eq:Epsilon_Bt}) and  (\ref{eq:slip_period}),
we obtain
\begin{equation}
\Tppm = \frac{10^4 P_0}{C \cos \alpha} \left(\frac{\Bt}{10^{16}{\rm G}}\right)^{-2}
\label{eq:T_Bt}
\end{equation}
where  $P_{\rm pr}$ is rewritten as  $P_0$ (in s), 
and $\Tppm$ is also in units of seconds.
This would enable us to estimate $\Bt$ from the PPM information,
but  two unknowns, $\alpha$ and $C$, are still left over.
At present, it is still difficult to deduce $\alpha$ from the data,
and what can be said is that neither $\alpha \approx 0$ (an aligned rotator)
nor $\alpha \approx 90^\circ$ (a flat spin case)  is likely,
because both these conditions would predict $A \approx 0$.
Likewise, $C$ must be   subject to  a fair amount of uncertainty,
arising from those in detailed magnetic-field distributions inside the stars,
and  in the nuclear equation of state.
Accordingly, we define a quantity
\begin{equation}
\Bt' \equiv \Bt \sqrt{ C \cos \alpha}
\label{eq:Bt'}
\end{equation}
and use it in place of $\Bt$. Then, Equation~(\ref{eq:T_Bt}) reduces to
\begin{equation}
\Bt'/(10^{16}{\rm G}) = \sqrt{(P_0/\Tppm)\times 10^4}.
\label{eq:Bt_final}
\end{equation}
The values of $\Bt'$ of the seven magnetars, calculated in this way,
are given in Table~\ref{tbl:PPM_summary},
together with their $\Bd$ and $\tauc$ from \cite{McGill}.

\begin{figure}
\centerline{
\includegraphics[width=8.5cm]{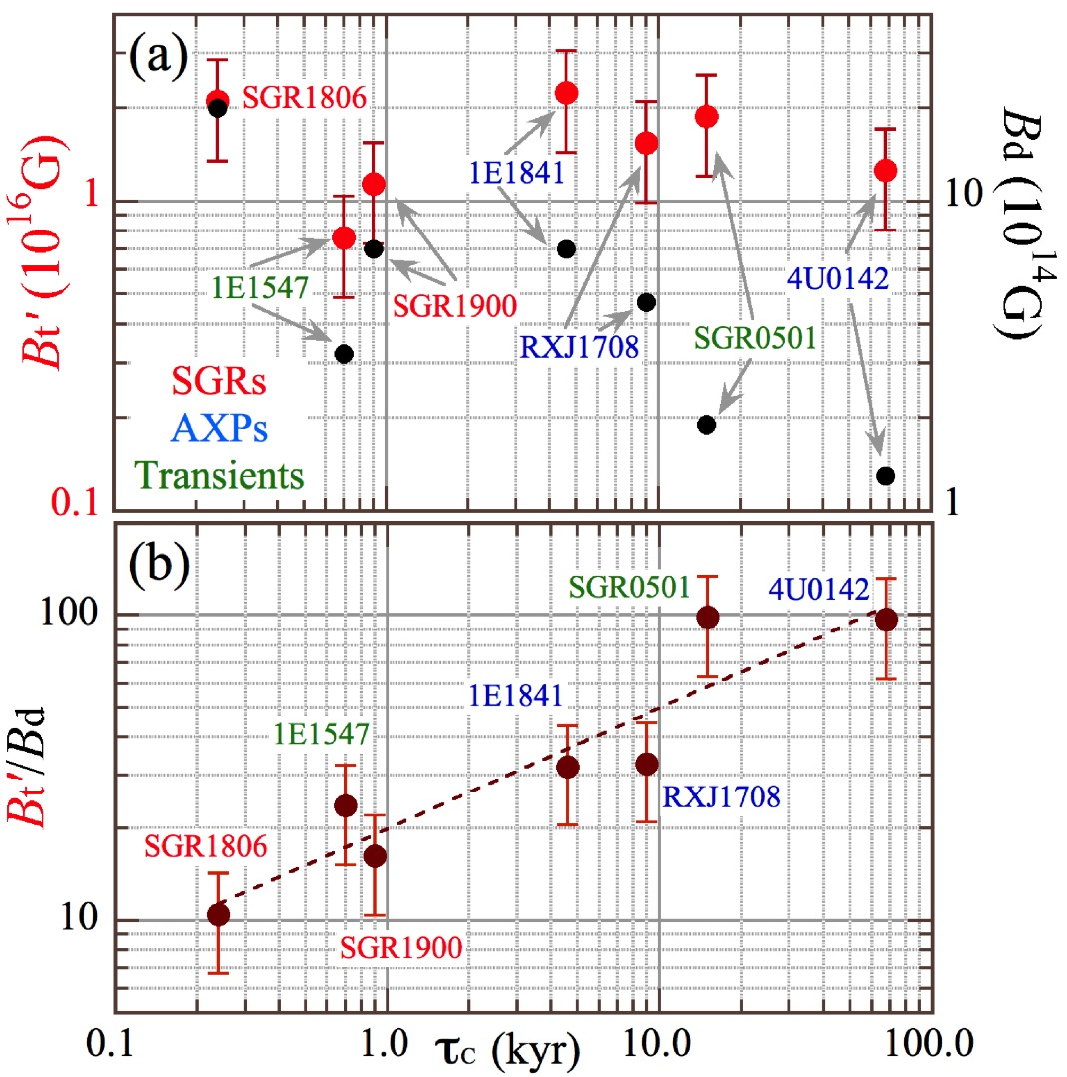}
}
\caption{
(a) Values of $\Bt'$ (red; left ordinate) defined in Equation~(\ref{eq:Bt'}),
and of $\Bd$ (black; right ordinate),
for the seven magnetars (Table~\ref{tbl:PPM_summary})
plotted against  $\tauc$.
See text for the estimation of the  errors in $\Bt'$.
(b) Their $\Bt'/\Bd$ ratios, shown as a function of $\tauc$.
}
\label{fig:MF_evol}
\end{figure}

Fig.~\ref{fig:MF_evol}(a) shows $\Bd$  and $\Bt'$  of these objects 
as a function of $\tauc$.
There, the error bar  in 
$\Bt'$ quadratically sums two error sources.
One is statistical errors in the  $\Tppm$ determination,
which overwhelms those in $P_0$.
The other is systematic uncertainty by a factor $\sim \sqrt{3}$
arising from $\cos \alpha$, 
assuming  that $\alpha$ varies from object to object,
but stays in a range between $20^\circ$ and $70^\circ$.
Because $C$  must  be more or less common to our sample objects,
we ignore its uncertainty which would not matter
as long as relative values of $\Bt$ witin the sample are concerned.
Thus, $\Bd$ exhibits  an indication of secular decay,
in agreement with the basic property of magnetars
that they are magnetically powered sources.
In contrast, $\Bt'$ shows no particular evolution.

\subsubsection{Evolution of the $\Bt'/\Bd$ ratio}

When  the toroidal-to-dipole component ratio, $\Bt'/\Bd$,
is plotted against $\tauc$, Fig.~\ref{fig:MF_evol}(b) is obtained.
This time,  a very clear evolution emerges
with a high correlation coefficient of 0.87,
in such a way that the ratio increases 
approximately as $\Bt'/\Bd \propto \tauc^{\beta}$
with $\beta \approx 0.4$.
A hypothesis of a constant $\Bt'/\Bd$ is ruled out.
Thus, the present result reinforces the suggestion by \cite{Makishima24},
that the toroidal fields of magnetars last longer than their poloidal dipole  fields.

In Fig.~\ref{fig:MF_evol},
the source names are  color coded;
SGRs in red, AXPs in blue, and transients in green.
The evolutionary trend appears common to
the three magnetar subclasses.
The difference between SGRs and AXPs would be
simply that the former objects are  younger and more active.
However,  the difference between the transient and non-transient magnetars 
may not be explained in that way,
because the two transients in our sample
differ largely in $\tauc$.

How does Fig.~\ref{fig:MF_evol}(b) compare
with  various theoretical works on the magnetic evolution of magnetars?
For example, \cite{Kojima21} predict that $\Bt$ should decay faster than $\Bd$,
but our result is opposite.
\cite{Vigano13} also performed detailed numerical studies,
but the  toroidal field they assumed has a dipolar distribution,
with the sign changing across the stellar equator.
The axial stellar deformation induced by such a magnetic configuration
should be much smaller, than in a mono-polar $\Bt$ distribution
that will explain our result more naturally.
Thus, the  comparison with the theoretical predictions remains 
inconclusive at present, and is left as a future work.

Putting aside such a comparison,
the characteristic age $\tauc$ requires some caution.
Generally, this observable gives a good measure of the true age $\tau_0$ of  a pulsar,
only when its initial spin period is much shorter than the present value,
and its spins down is driven by the magnetic dipole radiation 
{\em under a constant dipole field.}
Obviously,  the latter condition does not hold for magnetars,
considering the magnetic-field decay as in Fig.~\ref{fig:MF_evol}(a).
Then, the face-value $\tauc$ calculated ignoring the field decay
will significantly overestimate $\tau_0$ \citep{Nakano15},
and $\beta$ would increase  
(e.g., $\beta \approx 0.6$) if $\tauc$ is replaced with $\tau_0$.
Furthermore, the  $\tau_0$ vs.  $\tauc$ relation  could be diverse,
depending on thermal and spin histories of the NS \citep{Vigano13}.
Nevertheless, it is assuring that the values of $\tauc$ of our sample objects
stay within a factor of a few of their $\tau_0$ that are estimated by  \cite{Pons&Vigano}
in their figure 20 using a detailed numerical model of 
thermal and rotational evolution of isolated NS.

\subsubsection{Some thoughts on the toroidal magnetic field}

In Fig.~\ref{fig:MF_evol}(a), the seven values of $\Bt'$ show
a rather small scatter, by about a factor of 3.0 (maximum-to-minimum).
Therefore, the scenario of magnetar formation is suggested
to involve a relatively well organized process of toroidal-field generation
that is not much affected by various  attributes of their progenitors.
In addition, $\alpha$ may be relatively similar among the seven sources.
To our greater surprise,  
the $\Bt'/\Bd$ ratios in Fig.~\ref{fig:MF_evol}(b) exhibit an even smaller scatter,
by a factor of about 2.5, around the power-law fit with 
$\beta=0.4$. Then, the dipole  magnetic fields of magnetars  
could be a small fraction of their toroidal fields 
that leaked out of the stellar surface.
The overall magnetic configuration may be as illustrated  
by a numerical work of \cite{Braithwaite}.

An intriguing question is whether the ``magnetic energy" 
powering these magnetars comes from $\Bt$, or $\Bd$, or both.
On one hand, the decay of $\Bd$ observed in Fig.~\ref{fig:MF_evol}(a) suggests
that $\Bd$ is contributing to the magnetar activity,
at least while they are younger than $\sim 100$ kyr.
On the other hand, the sporadic activities observed from the 
aged magnetar populations (see next subsection)
 must be powered by  $\Bt$, 
because their  $\Bd$ is not strong enough any longer.
Therefore, the energy source is suggested to change gradually
from the dipole to the toroidal components;
$\Bt$  is more plenty, but may be released more slowly.
Some clues might be obtained by comparing
activity patterns between the typical magnetars
and the presumably more aged objects.

The theoretical works on the NS magnetism,
such as cited above,
are mostly based on the fundamental assumption
that the magnetic field results from electrical currents 
(subject to the Hall drift) flowing inside the star.
An interesting alternative, though much less popular,
tries to interpret the phenomenon in terms of ferromagnetism \citep{Makishima16PJA},
caused by spin-polarization in the crust-region neutrons \citep{NeutronPolarization} 
or in the residual electrons \citep{Ohnish&Yamamoto}.
This alternative may  also be considered,
when trying to give theoretical explanations to the present results.

\subsubsection{Populations of aged magnetars}

The suggested longer sustainment of $\Bt'$ (and hence of $\Bt$)
has a straightforward implication.
It very naturally explains the presence of aged magnetar-like objects
that  apparently  have weak $\Bd$ but  supposedly  strong $\Bt$.
As  mentioned in Section \ref{sec:intro},
the candidates include so-called weak-field magnetars,
some types of CCOs, 
and several radio pulsars with very long periods up to $\sim 10^3$ s \citep{ULPM}.

Among the above candidates,
an object of particular interest  is 1E161348$-$5055,
the CCO in the supernova remnants RCW 103.
It  exhibits  a  6.67 h periodicity \citep{DeLuca06},
which  would be too long to be explained via the 
spin-down of such a magnetar as in our sample.
Instead, as pointed out by \cite{RCW103_precession},
this unusually long period might  be $\Tppm$,
rather than the  pulse period of the object.
When the PPM has a large amplitude (e.g., $A \sim P/4$),
the true pulsation would not be detected directly, 
whereas the emission intensity may vary periodically,
at $\Tppm$ or possibly at $\Tppm/2$ (see, e.g., figure 12 of \citealt{Makishima23}).
This is because the beaming axis of the objects's emission,
relative to our line-of-sight,
varies both with the pulse phase and the modulation phase.
Under favorable geometrical configurations,
the expected intensity variation at $\Tppm$
can be so large \citep{RCW103_precession}
as to explain the factor $\sim 3$ flux modulation
seen in 1E161348$-$5055 \citep{DeLuca06}.
However,  to be honest,
we have so far examined none of our datasets 
for actual {\em intensity} variations at $\Tppm$,
because this requires (unlike the timing studies)
accurate background subtraction.
Therefore, this idea is yet to be confirmed through observations.

Another related phenomenon is so-called FRB repeaters, 
including  FRB 180916.J0158+65 with a 16.35 d period \citep{CHIME},
and  FRB 121102 with a $\sim 147$ d period \citep{FRBrepeater2}.
After a suggestion by \cite{Zanazzi20},
let us  examine whether these long periods can be identified with  $\Tppm$.
As a magnetar gets older, its $P_0$ will increase (though slowly),
and $\Bt$ would decrease even though the trend is not yet clear in Fig.~\ref{fig:MF_evol}(a).
In addition, $\alpha$ will increase with time,
because a prolate rigid both with a constant $\vec L$
attains the minimum energy  at  $\alpha = 90^\circ$.
Hence  in Equation~(\ref{eq:T_Bt}), the three factors, 
$P_0$, $\Bt^{-2}$, and $\cos^{-1} \alpha$, will all evolve so as to make $\Tppm$ longer
(though on unknown time scales).
If we consider a putative aged magnetar, 
with $P_0 = 20$ s, $\Bt/\sqrt{C}=0.5 \times 10^{16}$ G, and $\alpha=75^\circ$,
Equation~(\ref{eq:T_Bt}) predicts $\Tppm \sim 3$ Ms =35 days.
This estimate is  consistent with figure 1 of \cite{Zanazzi20},
and may well  explain the 16.35 d period of RB 180916.J0158+65.
However,  the $\sim 147$ d period of FRB 121102 may be too long.
Therefore, some, if not all, of these FRB repeaters
could be aged magnetars under free precession.
When they repeatedly emit collimated FRBs,
the bursts may reach us only at a particular modulation phase
wherein the beam points toward us.

\subsection{The two types of timing perturbations}
\label{subsec:discuss_perturbations}

In the present study,
two types of second-order timing perturbations were  observed,
both on top of the PPM which is regarded as the first-order effect.
One is the energy-dependent  lag in the pulse phase $\phi$,
observed from \SGR\ over  25.2--35.9 keV (Section \ref{subsec:ana_0501}; Appendix B).
This is  the same effect as detected
with \NuS\ and \ASCA\ in energies $E \le \Ee=10$ keV
from the gamma-ray binary LS~5039  \citep{Makishima23},
which is likely to  harbor a magnetar \citep{Yoneda20}.
The \NuS\ and \ASCA\ data indicate  
hard-lag  and soft-lag trends, respectively.
Similar behavior was also observed from 
\oneE\  \citep{Makishima21a} and SGR~1900+14  \citep{Makishima21b},
although in these cases $\phi$ varied with energy
quadratically rather than linearly.

The  other perturbation is the EDPV phenomenon
detected from \RXJ\ (Section \ref{subsubsec:ana_1708_EDPV_correction}; Appendix C).
The derived EDPV parameters are given in Table~\ref{tbl:EDPV},
together with those previously observed  from \oneE\  with \NuS,
as well as  from SGR 1900+14 \citep{Makishima21a} with \Su\  and \NuS.
The energy dependence observed thus far is always characterized with $\Delta \psi >0$
(clockwise motion toward higher energies as  in Fig.~\ref{fig:RXJ1708_09_AWscan}).

These two perturbations differ in 
that the former introduces an energy dependence in the pulse phase $\phi$, 
whereas the latter in the modulation phase $\psi$.
Nevertheless, they have two points in common.
One is that both are rather specific to magnetars,
because the phase-lag effect is rare among accreting X-ray pulsars,
with an exception being GS 0834-430 \citep{Miyasaka13}.
Similarly, the EDPV effect has not been observed from accreting objects,
since they lack the PPM behavior.
The other point is that the two perturbations both take place
in a sharply defined energy interval between $\Es$ and $\Ee$;
such behavior cannot be explained in terms of 
ordinary processes of thermal X-ray emission.

Considering these, the two timing perturbations are 
likely to provide  clues to the physics of strong magnetic fields.
Along that prospect, some thoughts were already given
in \cite{Makishima21a} and \cite{Makishima23IAU},
but  the outcome remains far from  satisfactory.
We do not discuss this issue any further.

\begin{table}
\caption{A summary of the EDPV parameters.}
\label{tbl:EDPV}
\begin{footnotesize}
\begin{tabular}{lcccccc}
\hline 
Object  &   Year\,$^{a}$   &  $\Es$ (keV) & $\Ee$ (keV) & $\Delta \psi$ (deg)\\
\hline \hline 
RXJ1708   & 2009H & $26.5 \pm 2.7$   & $44^{+8}_{-5}$  &$390 \pm 45$ \\
RXJ1708  & 2010H & $31.0 \pm  2.4$  &  $41 \pm 6$     & $216 \pm 45$ \\
\hline 
1E1547\,$^{b}$   & 2016N  & (10)$^{c}$     &  $26.6 \pm 0.8$ & $66.5 \pm 2.5$ \\
SGR1900\,$^{d}$ & 2009H &$13.5 \pm 0.4$  & $15.6\pm 0.5$   &  $174 \pm 22$\\
SGR1900\,$^{d}$& 2016 N & $13.2 \pm 1.8$  &  $21.0\pm 2.6$  &  $160 \pm 15$\\
\hline 
\end{tabular}
\begin{itemize}
\setlength{\itemsep}{0mm}
\item[$^a$] ``H" means the \Su\ HXD, and ``N" \NuS.
\item[$^b$] Results on \oneE\ from \cite{Makishima21a}.
\item[$^c$] $\Es=10$ keV is fixed in the analysis. 
\item[$^d$]  Results on  SGR 1900+14 from \cite{Makishima21b}.
\end{itemize}
\end{footnotesize}
\end{table}

\section{Conclusions}
Using archival X-ray data taken with the \Su\ HXD,
we performed  timing studies of
three magnetars, \Kes, \SGR, and 1RXS J170849.0$-$400910.
The  pulsation in their spectral hard X-ray component,
at a period $P_0$,
was found to be phase modulated (the PPM effect),
with a modulation period of $\Tppm=23.4$, 16.4, and 46.5 ks, respectively,
and a typical amplitude of $\sim $ 0.1 pulse cycles.
Together with the previous four examples
(4U~0142+61, \oneE, SGR 1900+14, and SGR 1806$-$20),
this effect is now confirmed  in seven magnetars in total,
and is regarded as  a common property of the NSs of this class.

These timing  results are interpreted in terms of free precession,
in which the NS is axially deformed to an asphericity of 
$\epsilon \sim P_0/\Tppm \sim 10^{-4}$.
Ascribing the deformation to  magnetic stress,
these magnetars are all inferred to harbor  
ultra-strong toroidal magnetic fields of $\Bt \sim 10^{16}$ G.

When we compare $\Bd$ of the seven magnetars,
with their $\Bt$ estimated from $P_0$ and $\Tppm$,
a clear evolution emerges;
the $\Bt/\Bd$ ratio increases toward older objects.
Therefore, magnetars are thought to retain their toroidal magnetic fields 
for a longer time than their dipole magnetic fields.
This can naturally explain the presence of some classes of NSs,
including FRB repeaters in particular,
which are considered to harbor strong toroidal magnetic fields
even though their dipole magnetic fields are lower than those of typical magnetars.

The present results also suggest
that the magnetar formation scenario  involves
a well organized process of toroidal-field generation.
In addition, the dipole magnetic fields of magnetars could be a
small fraction of toroidal magnetic fields
that leaked out of the stellar surface.

\section*{Acknowledgements}
The present work was  supported in part by the JSPS
grant-in-aid (KAKENHI), numbers 21K03624 and 24H01612.

\section*{Data availability}
The \Su\ HXD data underlying this article are publicly  available 
in the JAXA/DARTS Data Archive Website, 
at https://darts.isas.jaxa.jp/astro/suzaku/

\section*{Appendix A: Behavior of the $<15$ keV signal of  \Kes}
The data of \Kes\ were taken early in the \Su\ mission life,
and under a favorable sun-angle condition.
Hence, the lower energy boundary was in fact close to 10 keV.
To study the cause of the  DeMD peak suppression
below 15 keV  (Section \ref{subsec:ana_1841}),
we compared the pulse properties in two energy intervals,
10--15 keV and 15--25 keV, 
which contain comparable number of background-subtracted signal photons.
As given in Fig.~\ref{fig:f_AppA_E1841}(a), 
the $m=2$ periodograms from these two energy bands,
without demodulation,
have a nearly identical height and a fully consistent period.
Nevertheless, their $m=4$ DeMDs, shown in panel (b), are much different;
the 15-25 DeMD clearly reveals the 23.4 ks peak, 
and another peak at about half that period,
but they are absent in the 10-15 keV DeMD.

The \Su\ XIS+HXD  spectrum of \Kes\  in its  $\nu F_\nu$ form 
becomes minimum at $\sim 9$ keV \citep{Morii10},
where the two spectral components are thought to cross over.
In addition, as shown by \citet{Makishima21b},
the presence of the soft X-ray component would suppress the DeMD peak below an energy 
which is $\sim 1.3$ times this $\nu F_\nu$-minimum energy.
Since this would correspond to $\sim 12$ keV in \Kes,
the absence of the 23.4 ks peak in the 10--15 keV DeMD 
is consistent with the exclusive association of the PPM with the hard X-ray component.

\begin{figure}
\centerline{
\includegraphics[width=7.8cm]{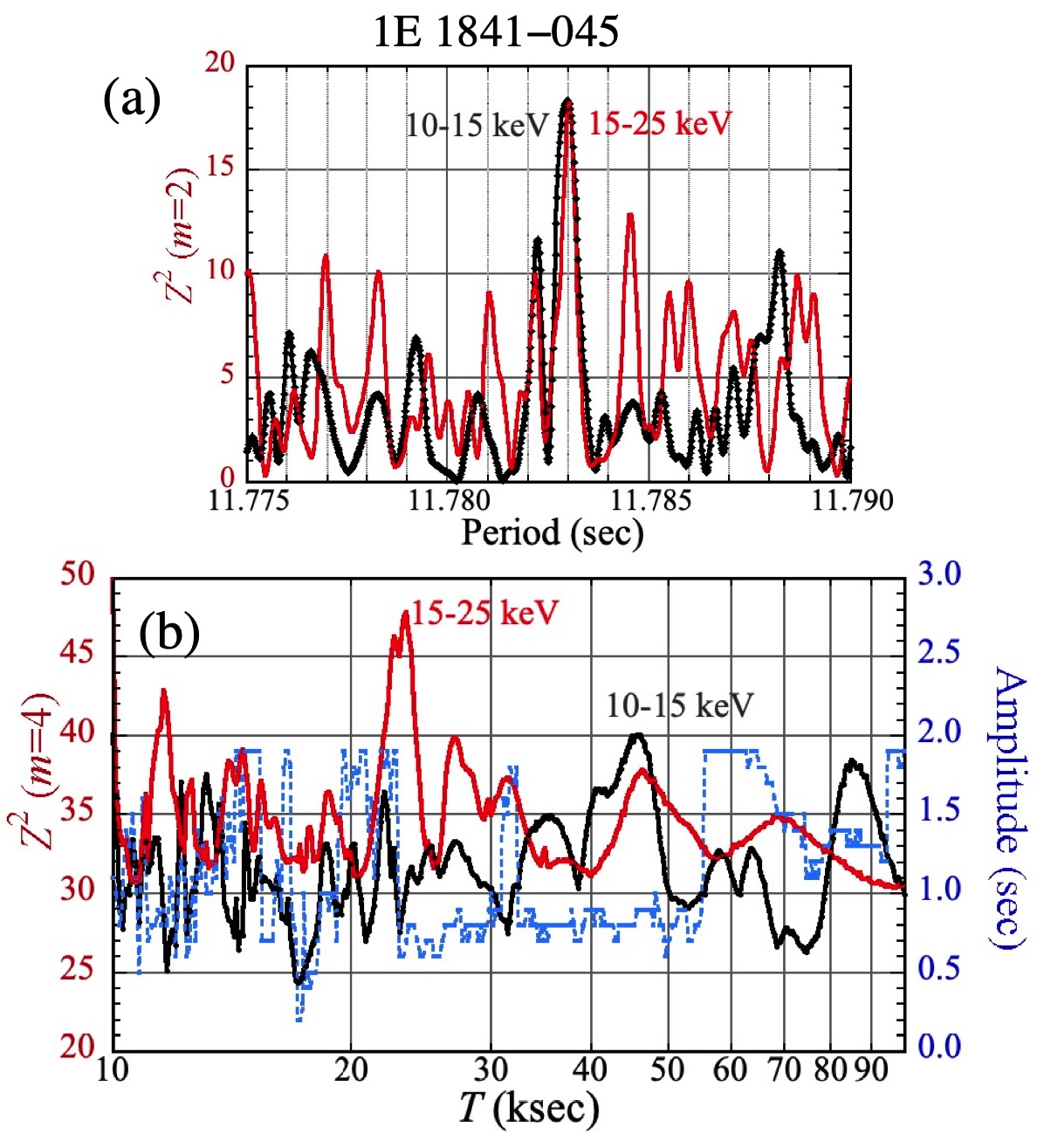}
}
\caption{
(a) Periodograms with $m=2$ of \Kes, in 10--15 keV (black) and 15--25 keV (red),
calculated in the same way as panel (b) of Fig.~\ref{fig:f1_1841}.
No demodulation is performed.
(b) DeMDs ($m=4$) of \Kes, in the same energy bands as panel (a).
The  blue line indicates $A$ in 10--15 keV.
}
\label{fig:f_AppA_E1841}
\end{figure}

\section*{Appendix B: Pulse phase lag in \SGR}

From $\sim 25$ keV up to $\sim  30$ keV, 
the pulses of SGR 0501+4516 are possibly affected by some  timing disturbance,
other than the PPM (Section \ref{subsec:ana_0501}).
We hence  produced folded pulse profiles in Fig.~\ref{fig:f_AppB_Pr},
in energy ranges below (red), right on (green), and above (blue)  the suspected interval.
Although the three profiles all reveal the pulsation,
they are not phase aligned.
Assuming that the pulse phase shifts with energy,
we can track the highest peak as indicated 
by dashed cyan (hard lag) or orange (soft lag) lines.
Manual phase adjustments referring to the peaks and valleys,
either hard-lag (panel b) or soft-lag (panel c) option,
brings the profiles into a much better phase alignment.
As a result, the profiles in black in (b) and (c),
obtained by adding up the three constituent profiles,
have larger amplitudes than that in (a).

To quantify the phase-lag behavior, 
let  $\phi_0~(0 \le \phi_0 \le 360^\circ)$ be the pulse phase of an event,
 determined by its arrival time and $P_0$.
Then, depending on the energy $E$ of that event,
we modify its pulse phase as \citep{Makishima23}
\begin{equation}
\begin{array}{lcc}  & \\
        \phi(E)  = &\\
                 & \end{array}  \left\{  
\begin{array}{ll}
 \phi_0    &(E \le \Es) \\
\phi_0 +\left( \frac{E - \Es}{\Ee-\Es}\right)\Delta \phi  & (\Es < E < \Ee)\\
\phi_0 + \Delta \phi                                                    & (\Ee < E)
\end{array} \right.
\label{eq:pulse_phase_lag}
\end{equation}
using three parameters; 
the start energy $\Es$,  the end energy $\Ee$,  
and the overall phase lag $\Delta \phi$.
Here, $\Delta \phi>0$ and $\Delta \phi<0$ mean
hard-lag and soft-lag conditions, respectively.

\begin{figure}
\centerline{
\includegraphics[width=5.5cm]{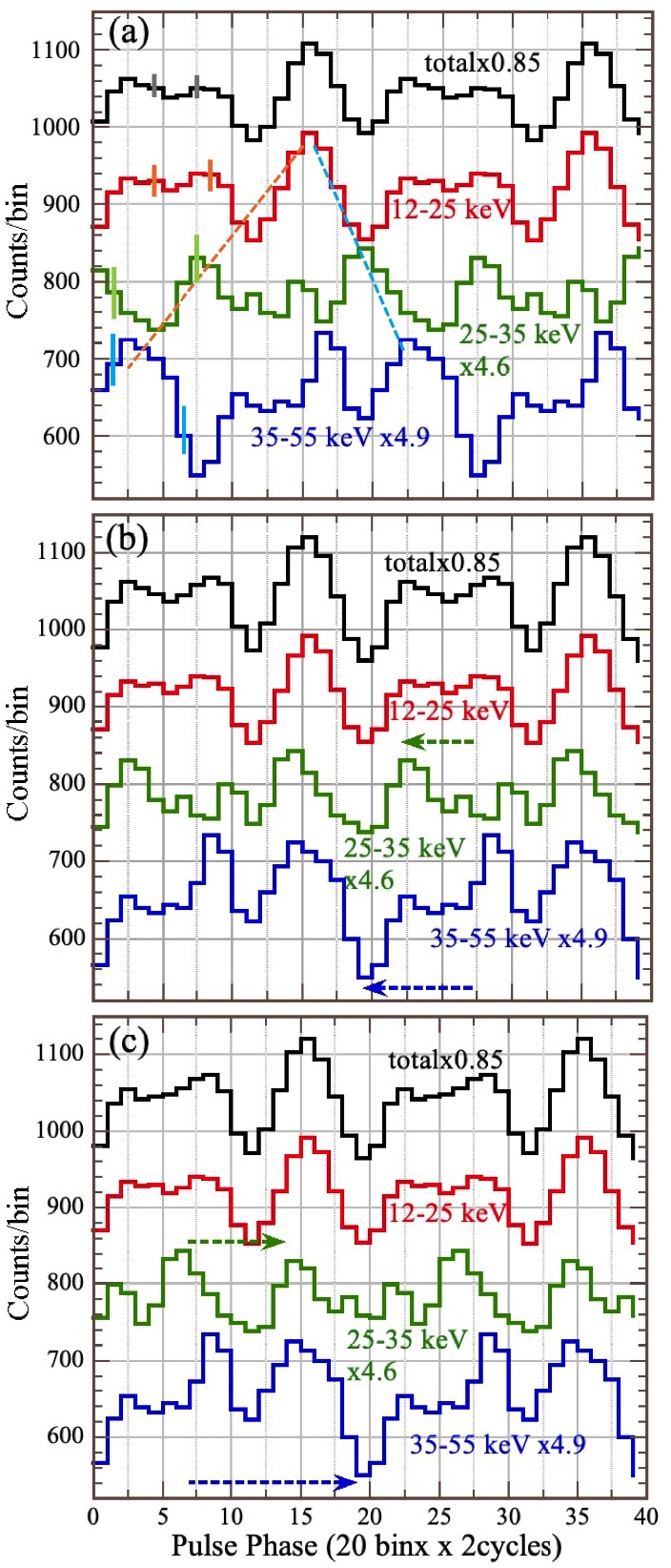}
}
\caption{
(a) Pulse profiles of \SGR, in 12--25 keV (red), 25--35 keV (green),
35--55 keV (blue), and their sum (black),
demodulated and folded using the same parameters in Table~\ref{tbl:results}.
For presentation, they are multiplied by appropriate scaling factors.
(b) The green and blue profiles are shifted to left
by 5 and 8 bins (see dashed arrows), respectively, relative to the red one.
This means a correction for a hard lag,
as indicated  in (a) by a cyan line.
(c) A soft-lag interpretation indicated by an orange line in (a).
The green and blue profiles are shifted to right by 7 and 11 bins, 
respectively, relative to the red one.
In (b) and (c), the profile in black sums
over the manually-adjusted three profiles.
}
\label{fig:f_AppB_Pr}%
\end{figure}


\begin{figure}
\centerline{
\includegraphics[width=7.5cm]{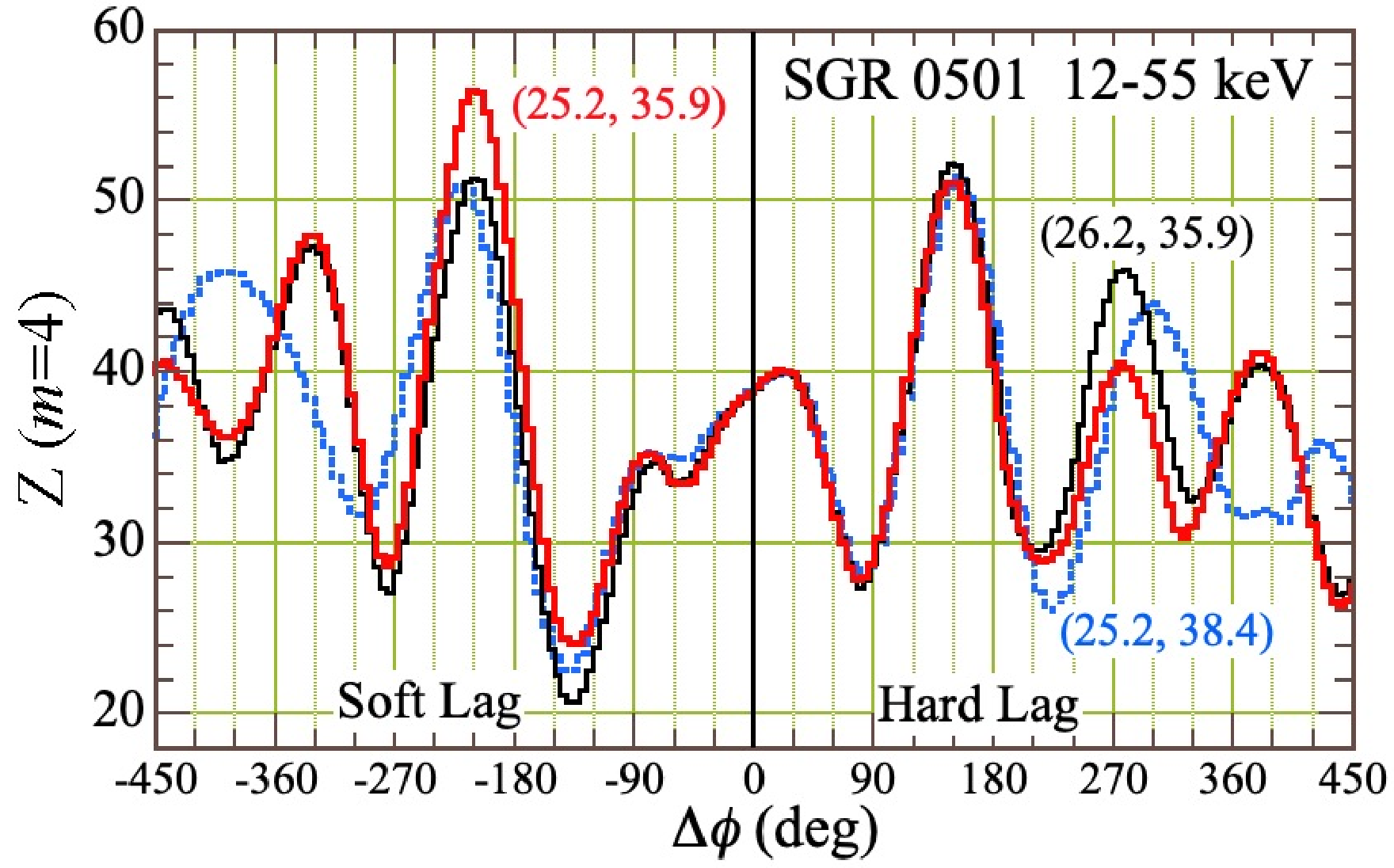}
}
\caption{
The maximum $\zz$ values in 12--55 keV from \SGR,
shown as a function of $\Delta \phi$ of Equation~(\ref{eq:pulse_phase_lag}).
At each $\Delta \phi$, $P$ is scanned over 
$5.76202 \pm 0.00006$ s in 6 steps,  
$A$ from 0 to 1.0 s ($\Delta A=0.1$), 
and $\psi$ from 0 to $360^\circ$ ($\Delta \psi=10^\circ$), 
but $T$ is fixed at $16.4$ ks.
The three curves employ different combinations of $(\Es, \Ee)$
as given in the figure.
}
\label{fig:f_AppB_Rscan}
\end{figure}

By correcting  the pulse phase of individual photons with Equation~(\ref{eq:pulse_phase_lag}),
we studied how the $T=16.4$ ks peak in the $m=4$ DeMD in 12--55 keV
varies with $\phi$, as it is scanned with a step of $5^\circ$.
The results are presented in Fig.~\ref{fig:f_AppB_Rscan}
for three typical combinations of $(\Es, \Ee)$.
We observe two peaks of comparable heights,
at  $\Delta \phi \approx 150^\circ$
and $\Delta \phi  \approx-210^\circ$,
corresponding to the  hard-lag and soft-lag interpretations 
in Fig.~\ref{fig:f_AppB_Pr}, respectively.
In addition, their values of $\Delta \phi$ just add up to become  $\approx 360^\circ$;
this explains why  the blue profile in  Fig.~\ref{fig:f_AppB_Pr}(b)
is identical to that in  (c).
However, when photons between $\Es$ and $\Ee$ are included,
the soft-lag peak becomes higher by $\sim 4.7$.
We hence adopt this option.
After further trimming the soft-lag parameters,
their optimum values were obtained as 
$\Es= 25.5\pm  1.0$ keV, $\Ee= 35.9^{+2.5}_{-3.7}$ keV,
and $\Delta \phi = -210^\circ \pm 10^\circ$.
Using these parameters,
we recalculated the entire DeMD,
and show it in Fig.~\ref{fig:f2_0501}(b) in orange.
The associated values of $P_0, T, A$, and $\psi$ are listed in Table~\ref{tbl:results}.
Thus, the correction with Equation~(\ref{eq:pulse_phase_lag})
considerably enhances the pulse significance.
The final 12--55 keV pulse profile derived in this way is very similar, 
though not identical, to the black profile in Fig.~\ref{fig:f_AppB_Pr}(c).

\section*{Appendix C: Modulation phase variation in \RXJ\ }

\begin{figure}
\centerline{
\includegraphics[width=6.5cm]{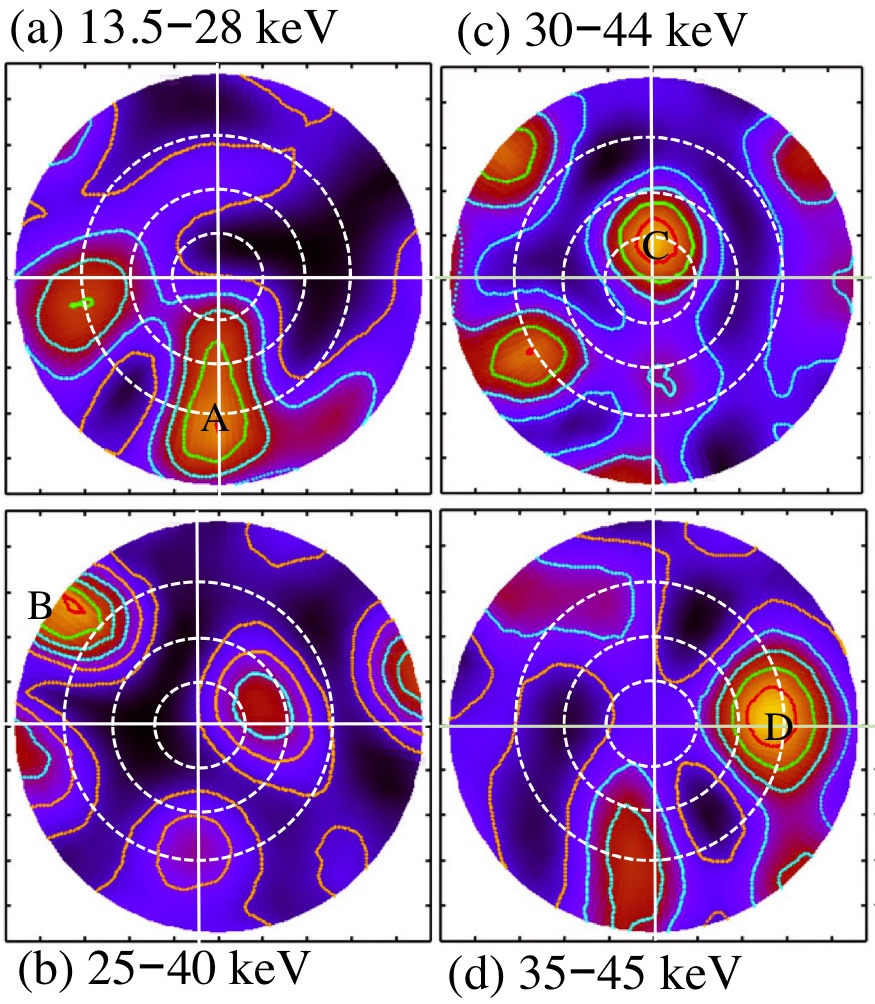}
}
\caption{
The same polar plot as Fig.~\ref{fig:f1_1841}(d1),
but from the 2009 data of \RXJ\ in four partially-overlapping energy bands.
The highest peak is indicated as A through D.
}
\label{fig:RXJ1708_09_AWscan}
\end{figure}

In the 2009 observation of \RXJ,
we encountered another type of timing disturbance
(Section \ref{subsubsec:ana_1708_EDPV_correction})
that is different from the pulse-phase lag observed from \SGR.
One possibility is the ``Energy Dependent Phase Variation (EDPV)" effect,
i.e., energy dependences in the modulation phase $\psi$,
sometimes accompanied by variations in $A$.
It was observed from \oneE\ \citep{Makishima21a} with \NuS,
and from SGR 1900+14 with \Su\ and \NuS\ \citep{Makishima21b}.
To examine this possibility, 
polar plots similar to Fig.~\ref{fig:f1_1841}(d1) 
were produced in various energy intervals,
assuming $P\approx P_0$ and $T\approx 46$ ks.
As presented in Fig.~\ref{fig:RXJ1708_09_AWscan},
the 46 ks DeMD peak clearly moves on the $(A,\psi)$ plane,
mainly clockwise toward higher energies,
accompanied by some changes in $A$ as well.

We tried to cancel the suggested EDPV effect,
by correcting the arrival times of individual photons.
This can be done using nearly the same formula as Equation~(\ref{eq:pulse_phase_lag}),
but replacing the pulse phase $\phi$ with the modulation phase $\psi$,  as
\begin{equation}
\begin{array}{lcc}  & \\
        \psi(E)  = &\\
                 & \end{array}  \left\{  
\begin{array}{ll}
 \psi_0    &(E \le \Es) \\
\psi_0 +\left( \frac{E - \Es}{\Ee-\Es}\right)\Delta \psi  & (\Es < E < \Ee)\\
\psi_0 + \Delta \psi                                                    & (\Ee < E).
\end{array} \right.
\label{eq:EDPV}
\end{equation}
A value of $\Delta \psi >0$ means 
that the DeMD peak on the $(A,\psi)$ plane
moves clockwise with the energy,
as in Fig.~\ref{fig:RXJ1708_09_AWscan}.

In the 2009 data of \RXJ,
the start energy is likely to be at $\Es \sim 27$ keV,
because the cyan peak in Fig.~\ref{fig:f3_1708}(c) 
decreased for $\EU >28$ keV.
We hence fixed $\Es=26.5$ keV tentatively,
and calculated how $\zz$ of the 46 ks DeMD peak in 13.5--55 keV 
depends on $\Delta \psi$,
for three representative values of $\Ee$.
As shown in Fig.~\ref{fig:f_AppC_Rscan},
the result reveals a clear $\zz$ enhancement at $\Delta \psi \approx 390^\circ$,
particularly when  $\Ee=44$ keV is chosen.
Through a further trimming, the three EDPV  parameters
have been determined as in  Table~\ref{tbl:EDPV}.
Thus, between 26.5 keV and 44 keV,
the modulation phase is found to changes by a full cycle.

The same EDPV effect,
but with somewhat different parameters,
was also observed from the 2010 data of \RXJ.
This is described in Section~\ref{subsubsec:ana_1708_EDPV_correction}.

\begin{figure}
\centerline{
\includegraphics[width=7.cm]{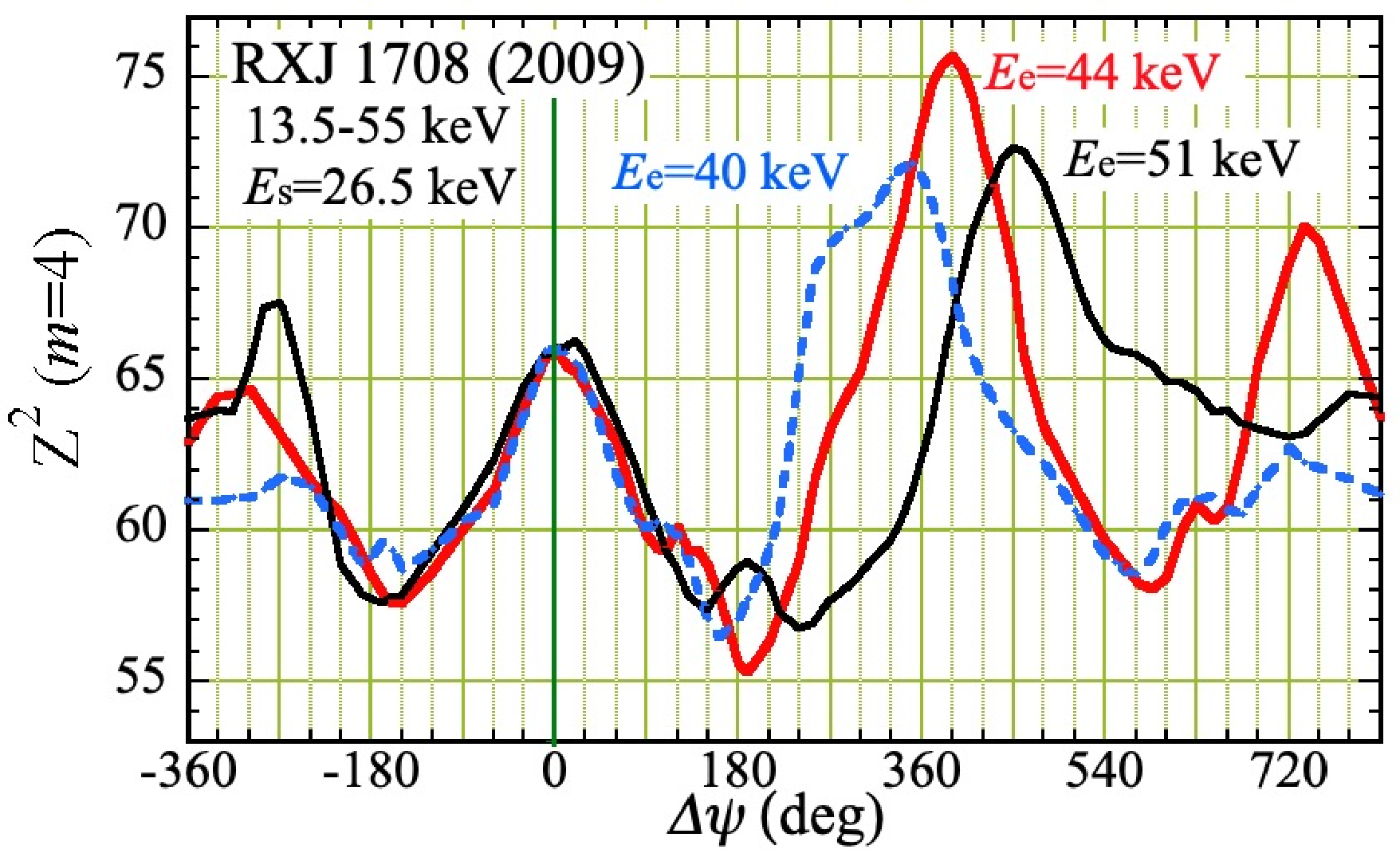}
}
\caption{
Dependence of the maxim $\zz$  on  $\Delta \psi$ of Equation~(\ref{eq:EDPV}),
in the  13.5--55 keV data of \RXJ\ in 2009.
As  in Fig.~\ref{fig:f_AppB_Rscan},
$P$, $A$, and $\psi_0$ are optimized at each  $\Delta \psi$, 
and $T$ is scanned over 43 to 47 ks with a step of 0.5 ks,
but  $\Es=26.5$ keV is fixed.
The three curves employ different values of $\Ee$ as given in the figure.
}
\label{fig:f_AppC_Rscan}
\end{figure}

\end{document}